\documentclass[12pt,preprint]{aastex}

\shorttitle{11 Mpc H$\alpha$ Survey}
\shortauthors{Kennicutt et al.}

\newcommand{\msun}{M$_\odot$}

\begin{document}


\title{Star Formation in NGC 5194 (M51a). II. \\
The Spatially-Resolved Star Formation Law}


\author{Robert C. Kennicutt, Jr.\altaffilmark{1,2,3}, 
Daniela Calzetti\altaffilmark{3,4,5}, Fabian Walter\altaffilmark{6},
George Helou\altaffilmark{7}, David J. Hollenbach\altaffilmark{8},
Lee Armus\altaffilmark{9}, George Bendo\altaffilmark{10}, 
Daniel A. Dale\altaffilmark{3,11}, Bruce T. Draine\altaffilmark{12},
Charles W. Engelbracht\altaffilmark{2,3},
Karl D. Gordon\altaffilmark{2,3}, 
Moire K.M. Prescott\altaffilmark{2}, 
Michael W. Regan\altaffilmark{5}, Michele D. Thornley\altaffilmark{13},
Caroline Bot\altaffilmark{9}, Elias Brinks\altaffilmark{14},
Erwin de Blok\altaffilmark{15}, Dulia de Mello\altaffilmark{16,17},
Martin Meyer\altaffilmark{5}, 
John Moustakas\altaffilmark{18}, Eric J. Murphy\altaffilmark{19},
Kartik Sheth\altaffilmark{9},
and J.D.T. Smith\altaffilmark{2,3}
}


\altaffiltext{1}{Institute of Astronomy, University of Cambridge, 
Madingley Road, Cambridge CB3 0HA, UK}
\altaffiltext{2}{Steward Observatory, University of Arizona, Tucson, AZ 85721}
\altaffiltext{3}{Visiting Astronomer, Kitt Peak National Observatory, 
National Optical Astronomy Observatory, which is operated by the 
Association of Universities for Research in Astronomy, Inc. (AURA) 
under cooperative agreement with the National Science Foundation.} 
\altaffiltext{4}{Department of Astronomy, University of Massachusetts,
710 N. Pleasant Street, Amherst, MA 01003}
\altaffiltext{5}{Space Telescope Science Institute, 3700 San Martin Drive,
Baltimore, MD  21218}
\altaffiltext{6}{Max-Planck-Institute f\"ur Astronomie, K\"onigstuhl 17, Heidelberg, D-69117, Germany}
\altaffiltext{7}{Caltech, MS 314-6, Pasadena, CA 91101}
\altaffiltext{8}{NASA Ames Research Center, MS 245-3, Moffett Field, CA 94035-1000}
\altaffiltext{9}{Spitzer Science Center, Caltech, MS 220-6, Pasadena, CA 91101}
\altaffiltext{10}{Astrophysics Group, Imperial College, Blackett Laboratory,
Prince Consort Road, London SW7 2AZ, UK}
\altaffiltext{11}{Department of Physics \& Astronomy, University of Wyoming,
Laramie, WY  82071}
\altaffiltext{12}{Department of Astrophysical Sciences, Princeton University,
Princeton, NJ 08544-1001}
\altaffiltext{13}{Department of Physics and Astronomy, 
Bucknell University, Lewisburg, PA 17837}
\altaffiltext{14}{School of Physics, Astronomy, and Mathematics, 
University of Hertfordshire, College Lane, Herts AL10 9AB, United Kingdom}
\altaffiltext{15}{Research School of Astronomy and Astrophysics,
Mount Stromlo Observatory, Cotter Road, Weston Creek ACT 2611, Australia}
\altaffiltext{16}{Laboratory for Observational Cosmology, Code 665, 
 Goddard Space Flight Center, Greenbelt, MD 20771}
\altaffiltext{17}{Department of Physics, Catholic University of
 America, Washington, DC 20064}
\altaffiltext{18}{Physics Department, New York University, 4 Washington 
 Place, New York, NY 10003}
\altaffiltext{19}{Department of Astronomy, Yale University, 
 P.O. Box 208101, New Haven, CT 06520-8101}


\begin{abstract}

We have studied the relationship between the star formation rate (SFR) 
surface density and gas surface density in the spiral galaxy M51a (NGC~5194),
using multi-wavelength data obtained as part of the Spitzer Infrared
Nearby Galaxies Survey (SINGS).  We introduce a new SFR index based
on a linear combination of H$\alpha$ emission-line and 24\,$\mu$m continuum
luminosities, that provides reliable extinction-corrected ionizing
fluxes and SFR densities over a wide range of dust attenuations.
The combination of these extinction-corrected SFR densities with 
aperture synthesis HI and CO maps has allowed us to probe the
form of the spatially-resolved star formation law on scales of
0.5 to 2~kpc.  
We find that the resolved  
SFR vs gas surface density relation is well represented 
by a Schmidt power law, which is similar in form and dispersion
to the disk-averaged Schmidt law.  We observe a comparably strong
correlation of the SFR surface density with the molecular gas 
surface density, but no significant correlation with the surface
density of atomic gas.  The best-fitting slope of the 
Schmidt law varies from $N$ = 1.37 to 1.56, with zeropoint and 
slope that change systematically with the spatial sampling scale.
We tentatively attribute these variations to the effects of 
areal sampling and averaging of a nonlinear intrinsic star formation
law.  Our data can also be fitted by an alternative parametrization
of the SFR surface density in terms of the ratio of gas surface density 
to local dynamical time, but with a considerable dispersion.  

\end{abstract}






\section{Introduction}

One of the crucial missing links in our knowledge of
star formation and galaxy evolution is an understanding of the 
interplay between the star formation rate (SFR) in
galaxies and the underlying properties of the interstellar medium (ISM).
Despite the physical complexity of this relationship,
observations of galaxies on global scales reveal
a surprisingly tight correlation between the average SFR per unit
area and the mean surface density of cold gas, extending over several
orders of magnitude in gas surface density (e.g., Kennicutt 1998a, b,
and references therein).  The most widely used parametrization is the
power-law relation introduced by 
Schmidt (1959, 1963).  In this paper we study the surface density form
of this relation:

\begin{equation}
\Sigma_{SFR} = A~\Sigma{{_g}{^N}}
\end{equation}

\noindent
where $\Sigma_{SFR}$ and $\Sigma_g$ refer to the star formation and
total (molecular and atomic) hydrogen surface densities, respectively.  
When measured in units
of M$_\odot$~yr$^{-1}$~kpc$^{-2}$ for $\Sigma_{SFR}$ and 
M$_\odot$~pc$^{-2}$ for $\Sigma_g$ the disk-averaged
data compiled by Kennicutt (1998b)
are best fitted by a power law with slope $N = 1.4 \pm 0.15$ and 
zeropoint $A = 2.5 \pm 0.7 \times 10^{-4}$.  This parametrization
has proven to be very useful as an input scaling law for analytical and
numerical models of galaxy evolution (e.g., 
Kay et al. 2002).  Throughout this paper we shall deal exclusively
with measurements of the surface densities of star formation and
gas, even if for the sake of concise text we do not refer explicitly 
to ``surface" density in every instance.

Despite its widespread application, this global law offers little
insight into the underlying physical nature of star formation regulation.
A surface density power law with $N \sim 1.5$ is consistent with what
one would expect if the SFR is mainly driven by large-scale gravitational
instabilities in the disk (e.g., Elmegreen 2002).  However the disk-averaged 
star formation rates of galaxies are nearly as well fitted by prescriptions
in which the SFR surface density scales with the ratio of  
gas surface density to local dynamical time (e.g., Silk 1997, Kennicutt
1998b):  

\begin{equation}
\Sigma_{SFR} = A^\prime~\Sigma_g~\Omega_g 
\end{equation}

\noindent
where $\Omega_g$ is the average angular frequency of the
rotating gas disk.  Such a relation could be
reproduced in a picture in which the SFR per unit gas mass in 
clouds is constant, and 
the frequency of cloud formation events scales inversely
with the local orbit time, for example the frequency of spiral 
density wave passages.  These scenarios are only cited as illustrative
examples; the fact that such different pictures can account for the
global SFRs underscores the inability of these data alone to discriminate
between different physical origins for the star formation law.

A complementary and in some ways more fundamental approach is to
use spatially-resolved measurements of the SFR and gas surface densities to 
examine the correlations between the observables on a point-by-point basis
within galaxies.  This allows us to quantify the form of the 
star formation law across the large ranges in local physical conditions
that are found in galaxies.
Numerous such studies of the local Schmidt law have been carried out 
over the last 40 years (see Kennicutt 1997 for a review).  Due to the
limited spatial resolution of the HI and CO data at the time, 
most studies were confined to the nearest galaxies, and correlated 
the SFR with either the atomic or molecular gas surface densities, but rarely
both.  The SFRs themselves were generally measured using blue star counts,
HII region counts, or H$\alpha$ emission, usually without corrections for
extinction.  Consequently it is not entirely surprising that the results of 
these studies have been inconsistent, with the derived power-law slopes 
(eq. [1]) ranging over $N = 1 - 3$ and beyond (Kennicutt 1997).  
More recently a number of workers have used radial profiles of gas
and SFR surface densities to constrain the form of the Schmidt law
on intermediate (typically few kpc) scales (e.g., Martin \& Kennicutt 2001,
Boissier et al. 2003, Schuster et al. 2007), and again the resulting
power law slopes are sensitive to the gas, SFR tracers, and the prescriptions
used to correct for extinction and convert the observed CO line intensities
into molecular gas surface densities.

The main limiting factors for this work have been the lack of  
high spatial resolution HI and CO observations of galaxies and 
of multi-wavelength observations of the star-forming regions,
which are necessary to derive accurate extinction-corrected SFR distributions.
The situation has improved in recent years with the completion of several
aperture synthesis CO mapping surveys of galaxies, most notably the 
Berkeley Illinois Maryland Association Survey of Nearby Galaxies 
(BIMA SONG; Helfer et al. 2003).  These provide
CO maps with synthesized beam sizes of several arcseconds,
making it possible to probe the form of the star formation law on 
sub-kiloparsec scales (e.g., Wong \& Blitz 2002).  However, dust extinction
poses a serious obstacle to these studies.  At the high gas column densities
probed by SONG, the corresponding dust column densities 
are large, producing attenuations of up 5 mag (or 
more) in H$\alpha$ and the ultraviolet.  This is large enough to cause
SFRs based on H$\alpha$ or ultraviolet measurements to be severely
underestimated.  On the other hand, much of the star formation
in disks occurs in regions with low to moderate extinction ($<$1
mag at H$\alpha$), and there estimates of SFRs based solely on the
dust emission will also be underestimated.  Moreover, since the extinction 
tends to correlate with the gas surface density itself, dust will
bias the slope of the derived star formation law if left uncorrected.
Wong \& Blitz (2002) used the gas surface density itself to estimate the
magnitude of the H$\alpha$ extinction correction, but as they pointed out 
this introduces a circularity into the determination of the star formation 
law, and it would be preferable to measure the attenuation corrections
independently of the gas density.

The advent of high-resolution infrared mapping of nearby galaxies
with the Spitzer Space Telescope now allows us to undertake a much
more rigorous study of the spatially-resolved Schmidt law.  The combination
of far-infrared, H$\alpha$, and Pa$\alpha$ maps of galaxies allows us
to derive extinction-corrected SFR distributions independently from the
CO and HI maps, and study the SFR vs gas surface density relation directly.
This investigation is one of the core science components of the Spitzer
Infrared Nearby Galaxies Survey (SINGS; Kennicutt et al. 2003).
The SINGS sample of 75 galaxies includes 24 objects
from the BIMA SONG survey, and comparable resolution HI maps have been
obtained at the Very Large Array (VLA) for a subset of SINGS spiral 
galaxies within 10 Mpc, as part of The HI Nearby Galaxy
Survey (THINGS; Walter et al. 2005).  The long-term goal
is to combine SFR distributions of these galaxies derived from SINGS
infrared and H$\alpha$ imaging with the CO and HI maps to study the
behavior of the star formation law down to scales of 6\arcsec, the
resolution of the Spitzer 24~\micron, BIMA, and THINGS maps.  This
corresponds to linear scales of 0.1 -- 0.5 kpc for most of the galaxies
in the subsample.

In this paper we present the results of a pilot study of the 
spatially-resolved star formation law in the nearby spiral M51a  
(NGC~5194).  This galaxy is ideal for a first study.  It possesses
a dense molecular disk with a high SFR per unit area,
and a large variation in extinction (A$_V$ $\sim$ 1 $-$ 4 mag),
which allows us to test the efficacy of our extinction
correction schemes.  In addition, the galaxy has an especially rich
multi-wavelength data set, including maps of the central disk in 
Pa$\alpha$ (Scoville et al. 2001), which can be used to derive 
extinction-corrected 
local SFRs independently of the {\it Spitzer} observations, and thus
quantify more accurately the uncertainties in the derived SFRs.
The SINGS observations of this galaxy have been presented in
an earlier paper (Calzetti et al. 2005; hereafter denoted Paper I).
Following that paper we adopt a distance of 8.2 Mpc to M51.

The remainder of this paper is organized as follows.  In \S 2 we 
describe the infrared, H$\alpha$, Pa$\alpha$, HI, and CO data that
were used for this study, and in \S 3 we describe the methods used
to extract local SFR and gas density measurements.  Over much of
the disk of M51 the intermediate levels of optical extinction introduce
large errors into SFR measurements based either on H$\alpha$ or infrared
fluxes alone, and a combination of measurements is needed to provide
reliable extinction-corrected SFRs.  In \S 4 we describe a new method
that we have devised to address this problem.  In \S 5 we present the
resulting SFR vs gas density relations, on varying linear scales and
for the total gas density as well as for the atomic and molecular 
components considered individually.  We also compare the spatially-resolved
relation in M51 to the global SFR law found for galaxies in general.
Finally in \S 6 we compare our results to theoretical expectations and
explore their implications for modeling star formation in galaxies.

\section{Data}

The primary derived parameters for this study of the star
formation law are local measurements of the SFR surface density 
and the atomic and
molecular gas surface densities.  We are interested in the 
instantaneous ($\tau <$ 5 Myr) star formation, and have employed
three tracers:  H$\alpha$ (0.66~\micron) and Pa$\alpha$ (1.87~\micron) 
imaging to measure the ionization rate, and the 24\,\micron\ dust
continuum imaging to trace the dust-obscured component of the 
star formation.  Combinations of these are used
to derive extinction-corrrected estimates of the SFR distribution.
We have used a combination of CO maps from BIMA SONG and other
published sources, along with 21-cm HI maps from the THINGS project
to map the surface densities of molecular and atomic hydrogen, respectively.
In this section we describe each of these data sets.

\subsection{Spitzer Infrared Images}

Spitzer MIPS observations of M51a at 24\,$\mu$m were
obtained on 22 and 23 June 2004, as part of the SINGS Legacy Program
(Kennicutt et al. 2003).  For this analysis we used the processed
MIPS images from SINGS Data Release 
4.\footnote{http://data.spitzer.caltech.edu/popular/sings} 
The reduction and mosaicing steps are described in Gordon et al. (2005)
and Bendo et al. (2006).  The final image mosaics have sizes
27$^{\prime}\times$60$^{\prime}$, fully covering M51a and the
surrounding background.  The 24\,$\mu$m image 
traces the thermal dust emission from the galaxy.  
In Paper I we carried out a detailed comparison of the star formation
in M51a as traced in Pa$\alpha$, 24\,\micron, and the ultraviolet, and
we refer the reader to that paper for a detailed discussion of the datasets.
In particular, Paper I revealed a very tight linear
correlation between the P$\alpha$-derived ionizing fluxes of the HII
regions in M51a and their 24\,\micron\ luminosities.  
In this paper we restrict most of our analysis of the Schmidt
law to discrete infrared and emission-line sources, and consequently
we will use the 24\,\micron\ fluxes exclusively as an infrared SFR tracer.
Imaging with MIPS at 70 and 160~$\mu$m was obtained as part of
the same observing campaign, but because of the low spatial resolution of those
data ($\sim$18\arcsec\ and 45\arcsec\ FWHM, respectively) they were 
not used in this project.   

The 24\,\micron\ map used for this paper is shown in Figure 1.  It has 
a diffraction-limited resolution of 5\farcs7 FWHM and a 1-$\sigma$
sensitivity limit of $1.1 \times 10^{-6}$ Jy~arcsec$^{-2}$ for isolated
sources.  The point spread function (PSF) displays prominent Airy diffraction
rings that limit the useful aperture sizes to approximately twice the
FWHM beam width.  The accuracy of the MIPS 24\,$\mu$m photometric 
zeropoint is $\pm$5\%\ (Engelbracht et al.\, 2007), although as discussed
later, other factors limit the accuracy of most of our aperture
fluxes to roughly $\pm$10\% (also see Paper I).

\subsection{H$\alpha$ Emission-Line Images}

Narrowband images centered at H$\alpha$ and continuum R--band images
were obtained on 28 March 2001, with the Cassegrain Focus CCD Imager on
the 2.1--m telescope at Kitt Peak National Observatory, 
as part of the SINGS ancillary data program (Kennicutt et al. 2003).
Two sets of exposures were taken to include
the entire extents of M51a and its companion NGC~5195 in the images.  
Exposure times were 1800~s and 360~s per position for
H$\alpha$ and R, respectively. Standard reduction procedures were
applied to the images.

Emission--line--only images were obtained by rescaling the R--band
image and subtracting it from the narrow--band image. 
The narrowband filter used for the observation contains contributions
from H$\alpha$ as well as the neighboring [\ion{N}{2}]$\lambda\lambda$6548,6583
forbidden lines.  M51a has been the subject of numerous
spectroscopic campaigns, and these data show an average [NII] excitation
that is near the asymptotic value for metal-rich HII regions 
of [NII]$\lambda$6548,6583/H$\alpha\ = 0.5$ (e.g., Bresolin et al. 1999,
2004).  We used this value to scale the image to net H$\alpha$ 
surface brightness.  In reality the 
[NII]/H$\alpha$ ratio varies somewhat from region to region and as
a function of radius in the galaxy, which introduces net flux errors
across the image.  Within the region covered by our CO data 
these variations introduce errors of $\sim$10\%\ or less
for most regions, and perhaps 20\%\ in the most discrepant cases.  

The final reduced H$\alpha$ image used is shown in Figures 1 and 2.
The accuracy of the absolute photometry 
was verified with an HST/WFPC2 H$\alpha$
image of the center of the galaxy.
The 1~$\sigma$ sensitivity limit of our final H$\alpha$ image is
1.8$\times$10$^{-17}$~erg~s$^{-1}$~cm$^{-2}$~arcsec$^{-2}$. The
measured PSF is 1.9$^{\prime\prime}$.

\subsection{Pa$\alpha$ Emission-Line Images}

The Pa$\alpha$ hydrogen recombination line at 1.87~$\mu$m provides a 
powerful probe of massive star formation even in relatively
highly obscured regions (Quillen \& Yukita 2001, Scoville et al. 2001). 
Likewise, the ratio of Pa$\alpha$ to H$\alpha$ flux provides a 
robust measurement of the nebular extinction.
An extinction of 1~mag at $V$ produces an
extinction of 0.15 magnitudes at Pa$\alpha$, i.e., a small, $\sim$14\%
change in the line intensity. We adopt an intrinsic flux ratio
H$\alpha$/Pa$\alpha$= 7.82 (Osterbrock \& Ferland 2006), 
which applies to an assumed electron temperature of 7000 K and 
electron density of 100 cm$^{-3}$ (Garnett et al. 2004, Bresolin et al. 2004).
Extinction corrections were derived using the extinction curve
of Cardelli et al.\, (1989), with 
k(Pa$\alpha$) = 0.455 and k(H$\alpha$)$-$k(Pa$\alpha$)=2.08,
where the extinction curve is expressed in the form
$I_{obs}(\lambda) = I_{intr}(\lambda) 10^{[-0.4*E(B-V)*k(\lambda)]}$.

Archival HST/NICMOS images of the central 144$^{\prime\prime}$, 
corresponding to the
inner $\sim$6~kpc of the galaxy, are available in the Pa$\alpha$
emission line (1.8756~$\mu$m, F187N narrow--band filter) and the
adjacent continuum (F190N narrow--band filter). The images form a
3$\times$3 NIC3 mosaic, and details of the observations, data
reduction, and image mosaicing are given in Scoville et al. (2001). The
nebular--emission--only image is obtained simply by subtracting the
F190N from the F187N image, after rescaling for the ratio of the
filter efficiencies. The NICMOS PSF is undersampled by the NIC3
0$^{\prime\prime}$.2 pixels, and the average 1~$\sigma$ sensitivity
limit of the continuum--subtracted image is
1.8$\times$10$^{-16}$~erg~s$^{-1}$~cm$^{-2}$~arcsec$^{-2}$,
which is approximately an order of magnitude less sensitive than the  
H$\alpha$ images, and another factor of eight lower when one considers
the lower intrinsic brightness of Pa$\alpha$ relative to H$\alpha$.
However this is partly compensated for by the much lower extinction
at Pa$\alpha$.  
Considering all these factors together
the effective sensitivity of the Pa$\alpha$ image to a fixed SFR per
unit area is roughly a factor of ten lower than that of the H$\alpha$
image, so its use is limited to the brightest HII regions in the inner 
disk.  Nevertheless it is critical for providing a calibration for the
other extinction correction methods used in this paper.

\subsection{Radio Continuum Fluxes}

Multi-frequency radio continuum fluxes are available for 43 bright
HII regions in M51a from the study of van der Hulst et al. (1988).
This study was based on aperture synthesis maps at wavelengths of
6~cm and 20~cm made with the Very Large Array (VLA), with matching
resolutions of 8\arcsec.  The two-frequency observations allowed these
authors to make a rough separation of non-thermal and thermal radio
fluxes.  The thermal bremsstrahlung luminosities provide 
independent measures of the  extinction-corrected
ionizing fluxes for the HII regions, and when combined with H$\alpha$
fluxes independent estimates of the visible extinction.  A full
description of these measurements can be found in 
van der Hulst et al. (1988).

\subsection{VLA HI Observations}

HI data for M51a have been obtained through The HI Nearby Galaxy
Survey (THINGS), a survey dedicated to obtain high-resolution VLA HI
imaging for $\sim$35 nearby galaxies (Walter et al.\, 2005). 
M51 was observed in the VLA D (9 July 2004), C (26 April
2004) and B array (05 March 2005) configurations for 80, 120, and 390
minutes on source, respectively (or a total of $\sim$ 10 hours on
source). The calibration and data reduction was performed using the {\sc
AIPS} package\footnote{The Astronomical Image Processing System ({\sc
AIPS}) has been developed by the NRAO.}.  The absolute flux scale 
was determined by observing 3C286 in all observing runs (using
the flux scale of Baars et al.\ 1977).  The same calibrator was
used to derive the bandpass correction.
The time variable phase and amplitude calibration
was performed using the nearby, secondary calibrators 1313+549 and 1252+565
which are unresolved for the arrays used.  The {\it uv} data were inspected
for each array, and bad data points due to either interference or
crosstalk between antennae were removed. 
After final editing and calibration, the data were combined to form a
single dataset and maps.

In order to remove the continuum we first
determined the line-free channels in our observation and 
subtracted the continuum emission in the {\it uv} plane. 
Datacubes (1024 $\times$ 1024 pixels $\times$ 80 channels each) were
produced using the task {\sc imagr} in {\sc AIPS}. To obtain
the best compromise between angular resolution and signal/noise, we used a
{\sc robust} parameter of 0.5 for the final imaging. This led to a
resolution of 5\farcs82 $\times$ 5\farcs56, and an rms of 
0.44 mJy\,beam$^{-1}$ in a 5.2 km\,s$^{-1}$ channel. 
The correspoding 3$\sigma$ sensitivity for the integrated map is 
$ 1.6 \times 10^{20}$ cm$^{-2}$ (corresponding to 1.3 
M$_\odot$\,pc$^{-2}$).  To separate real emission from noise
for the final integrated HI map (moment 0), we considered only those
regions which showed emission in three consecutive
channels above a set level ($\sim2\sigma$) in datacubes convolved
to 330\arcsec\ resolution.  The final corrected HI image is shown in Figure 1.

The fluxes in the integrated HI map are corrected for the fact that
typically the residual flux in cleaned channel maps is overestimated
(sometimes severely) due to the
different shapes of the dirty and cleaned beams (see e.g., 
Jorsater \& van Moorsel 1995, Walter \& Brinks 1999).  With these
corrections taken into account we estimate our column densities
to be correct within $\pm$10\%.

\subsection{BIMA SONG CO Observations}

The CO J=1--0 map of M51a was obtained as part of the BIMA Survey
Of Nearby Galaxies (BIMA SONG), and details on the data taking 
and processing can be found in Helfer et al. (2003). 
The map used in the analysis is shown in Figure 1.
The interferometric BIMA data on M51 were combined with single dish data
obtained at the former NRAO 12\,m telescope using on-the-fly
mapping. In total, a 26 pointing mosaic was observed with BIMA in the
C and D configurations, leading to a beamsize of 5\farcs8 $\times$
5\farcs1 (i.e., matched to the HI data) and a velocity
resolution of 4\,km\,s$^{-1}$ (similar to the HI
observations).  The map covers the 
central $\sim$350\arcsec, or $\sim$13.8~kpc of the galaxy.
The rms sensitivity in a 10\,km\,s$^{-1}$ channel is 61
mJy\,beam$^{-1}$.  The corresponding 3$\sigma$ sensitivity 
for the BIMA SONG map is $\sim$13 
M$_\odot$\,pc$^{-2}$ (Helfer et al. 2003).  This is
considerably higher than the corresponding HI surface
density limit; the limiting sensitivity of our cold
gas surface density measurements is set by the CO data.  

Molecular hydrogen column densities were calculated from the CO map,
using the conversion of Bloemen et al. (1986):
$ N(H_2) = 2.8 \times 10^{20}~I_{CO}$ cm$^{-2}$~({K~km~s$^{-1}$})$^{-1}$. The
high metallicity and small metallicity range in the disk of this galaxy
(Bresolin et al. 2004) justifies the use of a 
single conversion factor.  Our choice of the Bloemen et al. 
conversion factor is somewhat arbitrary and was done in part to maintain
consistency with the Schmidt law study of Kennicutt (1998b).  Moreover
this value lies in an intermediate range between lower factors
derived by Strong et al. (1988) and Hunter et al. (1997) and
higher values by Blitz et al. (2007) and Draine et al. (2007)
($1.56 - 4.0 \times 10^{20}$ cm$^{-2}$~({K~km~s$^{-1}$})$^{-1}$).  
As will be shown in \S 5.1, the choice of 
conversion factor mainly affects only the zeropoint of the Schmidt
law and not its form, because molecular gas is dominant over the
atomic component in the regions studied here, to such a degree
that the inferred total gas densities scale almost
linearly with the CO conversion factor used.

\subsection{FCRAO CO Data}

Single-dish CO maps covering the central 310\arcsec\ of M51a 
with a beam size of 
45\arcsec\ HPBW (1850 pc at the adopted distance) 
are also available from the study of Lord \& Young (1990).
The advantage of these data is that they provide fuller spatial sampling
of the disk (the sensitivity of the BIMA SONG maps only allows us
to measure CO intensities reliably in high surface brightness peaks), 
and thus they allow us to check whether the form of
the Schmidt law changes significantly when the spatial scales probed
are increased from 300 pc to 2 kpc.  As before, we have used the Bloemen et al.
(1986) conversion factor to calculate molecular hydrogen column
densities from the published CO fluxes.  

\subsection{Total Gas Densities}

With the conversion factors given above, molecular gas dominates
over the atomic component at most positions in the disk of M51a
(Scoville \& Young 1983, Lord \& Young 1990).  
However since most previous studies
of the Schmidt law have been parametrized in terms of the total
gas surface density we computed total hydrogen column densities
at most positions.  These in turn were converted to gas mass
surface densities assuming: 

\begin{equation}
\Sigma_H ~~( M_\odot~pc^{-2} ) = ((N_{HI} + 2N_{H_2}) / (1.25 \times 10^{20}~cm^{-2}).
\end{equation}

In order to maintain consistency with Kennicutt (1998b) we parametrized
the gas surface density in terms of hydrogen surface 
density alone.  Total gas surface densities,
including helium and metals, would be larger by a factor of 
$\sim$1.36, depending on the metallicity and dust depletion factor 
adopted.

\section{Measurement of Local SFRs and Gas Densities}

\subsection{General Considerations}

Before we describe the
local flux measurements and attempt to interpret the correlations we
find it is useful to examine qualitatively the 
behavior of the gas and star formation distributions.  These are illustrated in
Figure 1, which shows four of the maps that were used for the main part
of the analysis:  Spitzer 24~\micron, H$\alpha$\ (KPNO),
VLA HI, and BIMA SONG CO maps.  
%

Figure 1 shows that all of the gas and star formation tracers follow
qualitatively similar spatial distributions.  The strong tidal interaction
with NGC~5195 has organized the cold gas into 
strong and relatively narrow spiral arms, and the star formation
closely follows this structure.  Within the inner disk the gas
is predominantly ($>$90\%) molecular, with HI dominating only
outside of the main star-forming disk (Lord \& Young 1990).
This gas disk is dense, and the bulk
of the massive star formation is taking place in very massive
cloud complexes and giant HII regions.  When averaged over 
our primary aperture size (13\arcsec\ = 520 pc, see \S 3.2), typical cloud 
complexes have surface densities of order 10--1000~\msun~pc$^{-2}$
($N_H \sim 10^{22} - 10^{24}$ H~cm$^{-2}$), and  
gas masses within these apertures of order 
$2 \times 10^6 - 2 \times 10^8$~\msun.  Likewise the extinction-corrected
ionizing fluxes of the regions range over $\sim$10$^{49} - 10^{51}$
photons~s$^{-1}$, comparable to the 30 Doradus complex in the Large
Magellanic Cloud (e.g., Kennicutt 1984).  Consequently the nebular
ionization is provided by clusters and associations of order tens
to thousands of O-stars.  The unusual strength of the dynamical
disturbance and response of the gas disk in M51a offers both advantages
and disadvantages for this study.  Few nearby galaxies contain such
a large number of massive and dense star-forming complexes, and
this allows us to characterize the star formation law with more
than 200 regions.  On the other hand the combination of the strong
concentration of star formation in spiral arms with the finite 
sensitivity limits of our CO data restrict us to characterizing 
the behavior of the law in a high-density and high-SFR regime,
above the level where cloud formation and/or star formation
thresholds may become important.  

Generally speaking there is spatial correspondence between 
the locations of peaks in the cold gas components and star forming 
regions (see Figure 1d of Aalto et al. 1999).   
Upon close examination some subtle differences can be 
seen, such as a shift between maxima in the
CO and HI distributions, as discussed by Tilanus et al. (1988).
These displacements were interpreted by those authors as evidence 
that the HI is formed by photodissociation of the molecular gas by hot
stars located even farther downstream in the spiral pattern. 
Similar downstream offsets between the locations of the star formation peaks
(as traced at H$\alpha$, Pa$\alpha$, and 24\,$\mu$m) and the 
CO peaks are also observed at times (also see Aalto et al. 1999),
but a detailed analysis of this lies beyond the scope of the current study.
These displacements can be 
relevant to quantifying the form of the star formation law, by
imposing a limiting spatial scale over which we can make this comparison.
Fortunately the youngest star forming clusters traced in H$\alpha$, Pa$\alpha$,
and 24~$\mu$m have not drifted significantly
away from their natal clouds, and we can reasonably study the SFR density
vs gas density correlation on scales down to $\le$200 pc.  
However, as shown later the beam sizes and profiles
of our MIPS and CO/HI data impose a minimum aperture diameter of about
500 pc (12\farcs6 at the distance of M51a) anyway.  This scale 
is of prime interest for galaxy evolution modeling and
understanding the large-scale initiation and regulation of star formation.

When examined on these larger scales, we find an almost one-to-one
correlation between the locations of gas peaks and star formation events.
For example we were able to identify
only 2 significant CO peaks in the BIMA SONG map (out of 257 regions studied) that
do not show well-detected star formation in the visible and/or infrared.
The positions of these two clouds are highlighted by magenta circles
in an H$\alpha$ map shown in Figure 2.  Comparison with the CO map
in Figure 1 shows that they are prominent CO peaks, with surface
densities of $\sim$250 \msun ~pc$^{-2}$ or total molecular masses
of order $5 \times 10^7$~\msun, and are among the largest clouds in the
galaxy.  Both are located at the inside edge of the southern spiral arm,
which is at least qualitatively consistent with them being young clouds.
We find no significant emission features at these positions in
any of our other emission-line or Spitzer maps; examination of the
high-resolution HST images reveals the presence of a large dust
feature coincident with the southern cloud in the pair.  We 
tentatively conclude that these are very young clouds that have
yet to undergo significant massive star formation, though
we reiterate that such clouds are very rare in M51a, comprising
fewer than 1\%\ of the regions studied.   
This suggests a relatively short timescale for the onset 
of star formation relative to the lifetimes of the molecular concentrations,
at least under the (relatively extreme) conditions present in M51 today.
As a practical matter this result simplifies our analysis, because
it means that the form of the derived star formation law will not be 
sensitive to the method used to select the measuring apertures.

\subsection{Aperture Photometry and Local Flux Measurements}

Our analysis of the star formation law in M51a is based
on aperture photometry in Pa$\alpha$, H$\alpha$, and 
24~\micron\ (to determine local SFRs), and in CO and HI (for gas densities).
The primary dataset is based on 
measurements made with 13\arcsec\ diameter apertures, which
corresponds to a physical diameter of 520~pc at the
distance of M51. This aperture size is mainly dictated by the
point spread function of the {\it Spitzer} 24~\micron\ images.

Ideally one would measure the correlation between SFR and gas
surface densities by applying apertures at every position in 
the disk.  Unfortunately our data are not sufficiently sensitive
to permit this completely unbiased characterization of the
star formation law.  Because of the limited depth of the CO
maps, reliable gas surface densities can only be measured in
apertures that cover of order 10\%\ of the area of the star-forming
disk.  If we were
to ``measure" gas and SFR surface densities at every point in the
disk, $\sim$90\%\ of the data points would be upper limits in both
axes, and any correlations revealed by these data would be
physically meaningless.  With these limitations in mind we
adopted a two-part strategy.  To characterize the star formation
law on the smallest scales available to us we analyzed
13\arcsec\ aperture photometry restricted to 257 positions
(below) where star formation was detected (\S5.1).  Then to provide
a comparison set of (nearly) spatially complete measurements,
we used the 45\arcsec\ single-dish CO data of Lord \& Young (1990)
with matching infrared and H$\alpha$ aperture photometry (\S5.2).

The aperture positions for the 13\arcsec\ measurements are 
shown in Figure 2.  Centers were selected to coincide with 24~$\mu$m and
H$\alpha$ emission peaks, yielding a total of 257 positions.
In Figure~2 red circles indicate positions where significant
flux was detected in HI and CO, while blue circles indicate 
positions where we could only measure an upper limit in CO;
this is discussed in more detail in \S5.1.  The two magenta circles
mark the positions of CO concentrations without H$\alpha$ or 
infrared sources, as discussed earlier.

The photometric apertures were applied with a minimum separation of
centers of 7\farcs5 in the crowded central regions where Pa$\alpha$ 
was measured, and with minimum separations of 13\arcsec\ elsewhere.
This was imposed to minimize the
contamination from neighboring apertures in cases where the 24~$\mu$m
emission was used in conjunction with H$\alpha$ for deriving SFRs. In
particular, the MIPS 24~$\mu$m PSF gives contamination levels
of $\sim$5\% for an aperture centered 13$^{\prime\prime}$ away
(Calzetti et al. 2005).  The use of 13\arcsec\ apertures also
required the application of aperture corrections to the 24~\micron\ 
fluxes; we used a point-source correction factor of 1.67 
(Engelbracht et al. 2007).  Aperture corrections were not needed for 
data taken at other wavelengths.  For a few positions 
in the central disk the separation between sources is
less than the aperture diameter, producing overlapping
apertures; however only 8 of the 257 regions were affected by
this problem.

The need for relatively large measuring apertures introduces
considerable background contamination of the aperture fluxes in  
the H$\alpha$, Pa$\alpha$, and 24~$\mu$m measurements, from 
neighboring regions and diffuse background.  This 
contamination is strongest in the 24~\micron\ maps, which
contain an extended background component that is probably produced
in large part by dust heated by older stars (e.g., 
Popescu \& Tuffs 2002, Gordon et al. 2004).  This diffuse component
contributes 15--34\%\ of the total 24~$\mu$m luminosity of M51,
depending precisely on how the separation between point sources
and diffuse background is made (Paper I, Dale et al. 2007).
Conventional background
subtraction using annular regions around the apertures could
not be used here, because of contamination with neighboring regions
in many cases.  Instead we adopted the same strategy as used in Paper I 
for local background removal. We identified 12 rectangular areas,
each encompassing a
fraction of the 257 apertures, and fitted the local background
in each of these regions.  Figure 2 shows the locations of these 
background rectangles.  The net effect of subtracting backgrounds
is to reduce slightly the dispersion in the observed star formation
laws, but the best-fitting Schmidt laws are virtually identical
regardless of whether backgrounds are subtracted or not.
Background removal was not necessary for the HI and CO maps.

A similar process was used to measure 24\,$\mu$m, 
H$\alpha$ and HI fluxes for 45\arcsec\
apertures matching the published CO measurements of Lord \& Young (1990). 
In this case the positions of the apertures
were pre-determined by the CO measurements.  Background corrections
were applied to the H$\alpha$ and 24~$\mu$m data to
maintain consistency with analysis described above.

Measurement uncertainties assigned to the photometric values are a
quadratic sum of four contributions: random measurement
uncertainties in the raw source fluxes, variance of the local
background (from the original--pixel--size images), photometric
calibration uncertainties (5\%\ for Pa$\alpha$, H$\alpha$, and
MIPS 24~$\mu$m), and variations from potential mis-registration of the
multiwavelength images (at the level of 1$^{\prime\prime}$.5).
For the determination of gas surface densities the dominant
error term for most objects is measurement uncertainty in the CO flux.   
Conservative uncertainty estimates were employed on the CO map
to discriminate detections from upper limits. Such conservative estimates
were produced by quadratically combining the formal standard deviation with
the dispersion of multiple measurements obtained within a radius of 13
arcseconds (twice our fiducial measurements radius) in low signal--to--noise
regions. For the SFR
surface densities the relative uncertainties usually are
much lower, and are dominated by the source flux and/or
background and crowding terms.

The flux measurements in this paper are compiled in machine-readable
form in Table 1.  This includes positions for the 13\arcsec\ HII 
region measurements and the 45\arcsec\ areal photometry, for H$\alpha$,
Pa$\alpha$ (when available), 24\,$\mu$m, as well as HI and H$_2$ 
column densities derived from the radio maps.  We also compile the
derived, extinction-corrected SFR surface densities and total hydrogen
gas densities, as described in the next section.

\section{A New Combined H$\alpha\ +$ Infrared SFR Measure}

Any study of the spatially-resolved star formation in M51a must contend
with the substantial and locally variable dust extinction (e.g., Scoville
et al. 2001).  The median attenuation of the M51a HII regions is about
2 mag at H$\alpha$, with a range of 0--4 mag among individual objects 
(Paper I).
This means that neither H$\alpha$ nor 24~$\mu$m fluxes by themselves
provides reliable extinction-corrected SFRs across the disk.  Dust
extinction clearly will cause the H$\alpha$ fluxes to underestimate
the SFRs in most regions, often severely, while the infrared fluxes
by themselves will also underestimate the SFRs in the regions with 
low to moderate extinction.

In the center of the galaxy ($R \le$ 72\arcsec\ = 3 kpc) we have
Pa$\alpha$ imaging, and the combination of Pa$\alpha$ and H$\alpha$
photometry allows us to derive robust extinction corrections 
(the average extinction of the HII regions at the wavelength of 
Pa$\alpha$ is only $\sim$0.4 mag).  However Pa$\alpha$ data 
are not available for most of the disk of M51a, and even where
they are available they can only be applied to 77 high surface
brightness regions.  This effectively imposes a limiting SFR surface
density of $\sim$0.05 M$_\odot$~yr$^{-1}$~kpc$^{-2}$, 
only $\sim$10 times lower than the maximum observed SFR density,
and insufficient to reliably define the form of the Schmidt law.
Therefore in order to extend the range of SFR densities probed we need
a second extinction-corrected SFR measure that does not rely on
Pa$\alpha$ data.  

In principle it should be possible to combine the observed (extincted)
H$\alpha$ fluxes with the infrared fluxes to derive extinction-corrected
emission-line luminosities, because the infrared emission
comprises most of the stellar luminosity that was attenuated by the dust.
A variant of this approach was introduced by Gordon et al.
(2000) as the ``flux ratio method," in which they used the combination of
ultraviolet and infrared fluxes of galaxies to derive extinction-corrected
UV luminosities and SFRs (also see Bell 2003, Hirashita et al. 2003,
and Iglesias-Paramo et al. 2006).  Here we introduce a similar method,
but one which combines measurements at H$\alpha$ and 24~\micron\ to
derive extinction-corrected H$\alpha$ and ionizing luminosities.

The basis for this method is the observation in Paper I of a very tight
and linear correlation between 24~\micron\ and extinction-corrected 
Pa$\alpha$ luminosities for 42 dusty HII regions in the center of M51a
(where 80--90\%\ of the total stellar luminosity is reradiated in the
infrared).  Subsequent work has shown that this trend
extends to other highly-extincted HII regions in other galaxies 
(Wu et al. 2005; Alonso-Herrero et al. 2006; Calzetti et al. 2007).
This correlation allows us to empirically
calibrate a relation between 24~\micron\ luminosity and the SFR 
that is directly tied into the H$\alpha$ (or Pa$\alpha$) based scale.
A tight linear scaling between ionizing flux and 24~\micron\ flux
might be expected if single-photon heating of small dust grains
($<$50\,\AA\ in radius) dominates the emission in this wavelength range,
or if the average temperature of the emitting dust does not vary 
substantially from position to position. 

Following the precepts of Gordon et al. (2000) we can parametrize
the H$\alpha$ attenuation in terms of a simple energy balance.  To
a first approximation the amount of extincted H$\alpha$ radiation
should scale with the luminosity re-radiated in the infrared:

\begin{equation}
L(H\alpha)_{corr} =  L(H\alpha)_{obs} + {a~{L(24)}} 
\end{equation}

\noindent
where $L(H\alpha)_{obs}$ and $L(H\alpha)_{corr}$ refer to the observed
and attenuation-corrected H$\alpha$ luminosities, respectively, 
$L(24)$ is defined as the product $\nu L_\nu$ at 24~\micron, 
and $a$ is the scaling relation
that is fitted empirically, using independently extinction-corrected
data such as Pa$\alpha$ and H$\alpha$ measurements.  The
same relation can be used to measure the effective H$\alpha$ attenuation:

\begin{equation}
A(H\alpha) = -2.5 \log { {L(H\alpha)_{obs}} \over {L(H\alpha)_{corr}}}  
          = 2.5 \log { (1 + { {a~L(24)} \over {L(H\alpha)_{obs}}}) } 
\end{equation}

\noindent
Note that in the limit of zero extinction the infrared term vanishes,
and $L(H\alpha)_{corr} =  L(H\alpha)_{obs}$.  In
the opposite limit of very high extinction,  $L(H\alpha)_{obs}$ vanishes
and $L(H\alpha)_{corr} =  aL(24)$.  This defines the scaling constant
$a$.

The relations in eqs. (4$-$5) are empirical approximations to a
much more complicated extinction geometry in individual regions.
In a real HII region the ratio of observed H$\alpha$ luminosity 
to infrared luminosity will depend on the dust 
optical depths and geometry, which will influence the amounts of
extinction and the energy distribution of the emitting dust, and
on the spectral energy distributions of the embedded
stars, which affect the ratio of ionizing to dust heating radiation.
All of these factors will vary from object to object and introduce
a scatter into the relations between actual H$\alpha$ extinctions
and the values estimated from eq. (5).  Our interest is in
applying these relations statistically, and we can use the
observed scatter against independently determined luminosities
and extinctions to constrain the reliability of the results.

In M51a we have independent extinction-corrected H$\alpha$ luminosities
for 42 HII regions from the Pa$\alpha$ measurements, and we used these
to calibrate the mean value of $a$ in equation (4).  The results are
shown in the left panel of Figure 3.  There we compare the 
extinction-corrected H$\alpha$ luminosities of the HII regions derived
from the observed H$\alpha$ fluxes, 24~\micron\ fluxes, and eq. (4)
with extinction-corrected H$\alpha$ luminosities of the same objects
as derived from the ratio of Pa$\alpha$/H$\alpha$ (abscissa).  
We find a best fit for a L(24)/L(H$\alpha$) scaling constant
$a = 0.038 \pm 0.005$.  The rms dispersion of the individual regions
about the mean relation is $\pm$0.1 dex ($\pm$25\%), which provides
an empirical estimate of the accuracy of the attenuation corrected
luminosities.  The scatter reflects a combination of measuring uncertainties
in the Pa$\alpha$ and 24~\micron\ fluxes, along with errors in the 
application of equation (4) caused by variations in cluster age,
dust geometry, etc.  In the righthand panel of Figure 3 we compare
the corresponding $V$-band attenuations derived using the two methods.
The average scales are constrained to be the same, 
because we calibrated the coefficient in eq. (5) using these regions;
the main result of interest is the dispersion of points about 
the mean relation ($\pm$0.25 mag).  This can be compared 
to the systematic errors that would be introduced if no extinction
correction were applied, 1--3.5 mag (factor 2--25) for these regions.

We can also compare our extinction-correction values and luminosities
to those derived from a comparison of thermal radio continuum and 
H$\alpha$ fluxes by van der Hulst et al. (1988).  
Those authors used 6~cm and 20~cm
VLA maps of M51 to perform an approximate separation of thermal
(free-free) and non-thermal (synchrotron) components to the fluxes.
The thermal radio fluxes scale linearly with the ionizing fluxes,
with a mild dependence on electron temperature (assumed to be 
7000 K).  Estimated thermal radio fluxes at 6~cm are available for
32 HII regions in common with our sample.  The 
resulting luminosities for individual regions have larger uncertainties 
than those derived from Pa$\alpha$, because of the lower signal/noise 
of the radio data and uncertainties in the corrections for nonthermal 
emission, typically $\pm$20--50\%\ (van der Hulst et al. 1988).
However the data provide a valuable check on the overall extinction
and corrected flux scales.  The median radio-derived attenuation 
for the 32 HII regions is A(H$\alpha$) = 1.9 mag, which is similar to the
median value of 1.75 mag using eq. (5).  In view of the considerable
uncertainties in the radio data (typically $\pm$0.5 mag in derived
extinction at H$\alpha$) we regard this as reasonable consistency.
We defer further discussion of this method to a more extensive
analysis by Calzetti et al. (2007), which 
incorporates Pa$\alpha$, H$\alpha$, and 24~$\mu$m measurements 
of 220 HII regions in 33 galaxies, and reinforces the conclusions
drawn above.\footnote{The best fitting value of the calibration constant
derived in the Calzetti et al. analysis is slightly different, 
$a = 0.031$ {\it vs} 0.038 derived here.  For this analysis we have
opted to use the latter value, since it was derived from the same
data that are used to measure the SFRs in M51a.  However adopting the
other value of $a$ would not alter the results presented in this
paper significantly.}
In that paper we also show that the empirically
determined value of $a$ in eqs. (4$-$5) is consistent with
expectations from simple evolutionary synthesis models of young 
star clusters surrounded by gas and dust.

In the remainder of this paper we use equation (4) to estimate H$\alpha$
extinction corrections for the 215 HII regions in M51a that were not
measured in Pa$\alpha$ (and the 42 regions with Pa$\alpha$ data as well).
We hasten to emphasize however that our calibration of $a$ is based on
and tailored to the HII regions in M51a, and may not necessarily apply
in all physical situations.  In particular our determination of the
scaling factor $a$ for HII regions cannot be applied to galaxies as
a whole, because galaxies contain a significant component of 24~$\mu$m
dust emission that is not associated with HII regions.  The application
of this method to galaxies is addressed in a separate paper (Kennicutt
\& Moustakas 2007, in preparation).  Likewise one would expect the method
to break down badly in small HII regions that are predominantly ionized 
by single stars, because in such regions the ratio of ionizing luminosity
to dust-heating luminosity will be strong functions of ionizing stellar
type, age, and the cluster mass function.  These will vary enormously
(and systematically) from object to object.

\section{Results:  The Local SFR Density vs Gas Density Relation}

The measurements described in the previous section provided us with
extinction corrected emission-line fluxes for the 257 star-forming
regions in the area covered by the BIMA SONG map.  These include
Pa$\alpha$ measurements for 77 regions in the central 144\arcsec\
(corrected for dust attenuation via Pa$\alpha$/H$\alpha$) and
24~\micron\ + H$\alpha$ fluxes for 180 regions (these include 
25 objects in the inner 144\arcsec\ region that were not detected in 
Pa$\alpha$).

Up to now we have measured ionizing fluxes of HII regions and their
embedded OB associations, and we now would like to transform these
to equivalent SFRs and SFRs per unit area.  For HII regions with 
sufficiently high luminosity (($L_{corr}(H\alpha) \ge 10^{39}$
ergs~s$^{-1}$, ionizing photon flux $Q(H^0) \ge 10^{51}$~s$^{-1}$), 
the ionizing star
clusters need to be sufficiently massive such that their initial mass
functions will be well populated to high masses, and we can safely
assume that the ionizing fluxes (at fixed age) will scale roughly with the 
total stellar masses of the clusters (e.g., Kennicutt 1988,
Cervi\~no et al. 2002) and thus the SFR.  
With this in mind we have converted the line fluxes
into equivalent SFRs, using the calibration of Kennicutt (1998a)
that is usually applied to galaxies as a whole.

\begin{equation}
SFR ~(M_\odot ~yr^{-1}) = 7.9 \times 10^{-42} ~L_{corr}(H\alpha) ~~ (ergs~ s^{-1})
\end{equation}

\noindent
This conversion assumes a Salpeter IMF over the range of stellar
masses 0.1 $-$ 100 M$_\odot$.
We caution that a ``star formation rate" derived in this
way for an individual HII region, using a continuous star formation
conversion relevant to entire galaxies, has limited physical meaning,
because the stars are younger and the region under examination is
experiencing an instantaneous event when considered on any galactic
evolutionary or dynamical timescale.  One must also bear in mind that
age differences among the HII regions will change the actual ratio of
ionizing flux to stellar mass, and thus introduce scatter into the
derived Schmidt law.  However for this analysis we are mainly interested
in the shape of the star formation law, and the normalization of the
SFR scale is somewhat arbitrary.  Adopting a global conversion provides
a convenient standard and will also allow us 
to compare the form and zeropoint of the
relation with that measured for galaxies as a whole; this is discussed 
in the next section.

In order to cast these measurements in the form of a Schmidt law,
the SFRs then need to be converted to SFR surface densities by normalizing
the rates to an appropriate area.  We followed the
most straightforward approach of dividing the SFR by the projected
area of the 13\arcsec\ apertures, and divided by an
additional factor of 1.07 to correct for the 20\arcdeg\ inclination of M51a
(Tully 1974).  
The gas surface densities were corrected by the same projection factor.
This choice of normalization is somewhat arbitrary, 
but we believe it is the most
physically meaningful choice, because it corresponds to the approximate
size of the emitting regions, their associated gas complexes, and
to the sizes of the regions measured.  Note that the power law exponent
of the derived Schmidt law derived form is insensitive to the 
apertures used; adopting a larger aperture, for example, will simply 
decrease the measured gas and SFR surface densities by the same
beam dilution factor for most points.  This shift however will
change the zeropoint constant of the derived Schmidt law (discussed
in more detail in \S5.3).

\subsection{The Star Formation Law on 500 Parsec Scales}

One of the main results of our paper is summarized in Figure 4, which shows
the relationship between the SFR and gas surface densities for the 257 regions
covered by the BIMA SONG map, measured with apertures of 13\arcsec\
(520 pc) diameter.  Solid triangles denote SFRs measured from
Pa$\alpha$, while open triangles show those with SFRs determined from 
24~\micron\ and H$\alpha$ measurements.  
For the sake of clarity
we have removed the error bars in the right panel, while the
same data with error bars are shown in the left panel.  
Open circles in the righthand panel denote positions
where we only could determine an upper limit to the CO flux; for those
we plot the HI surface density as a lower limit and the sum of the
1$\sigma$ H$_2$ surface density plus the HI surface density as an upper limit.
The SFRs and gas surface densities
are strongly correlated, and follow a roughly power law relation
in the mean.  The solid line in both plots shows a bivariate least
square fit:

\begin{equation}
\log {\Sigma_{SFR}} = (1.56 \pm 0.04)~\log {\Sigma_H} - (4.32 \pm 0.09) 
\end{equation}

\noindent
where the SFR surface density $\Sigma_{SFR}$ is expressed in units of 
M$_\odot$\,yr$^{-1}$\,kpc$^{-2}$ and the hydrogen gas surface density
$\Sigma_H$ is expressed in units of M$_\odot$\,pc$^{-2}$.  
The uncertainties given in the equation refer to random fitting errors only.
Some of the data points shown in Figure 4 carry large uncertainty
estimates in the gas surface densities (left panel), and this may
give rise to concerns about the robustness of the fit given
above.  We tested this by refitting the data with 25 interarm
regions with large uncertainties in CO fluxes removed.  The
resulting relation 
($\log {\Sigma_{SFR}} = (1.57 \pm 0.05)~\log {\Sigma_H} - (4.36 \pm 0.09)$)
is the same within the formal errors, so this does not appear to
be a serious concern. 

The scatter in the correlation is significant,
with an rms dispersion about the best fit of $\pm$0.4 dex. 
This is comparable to the dispersion in the global Schmidt
law relation of Kennicutt (1998b).  
Fortunately M51a offers a large dynamic range in local SFR 
and gas surface densities (factors of roughly 1000 and 100, respectively),
so the correlation is well defined despite this large point-to-point
scatter.  

What are the likely sources of this dispersion, and does any of 
it reflect a real physical variation?  As indicated by the error bars
in the left panel of Figure 4, observational uncertainties in
the gas masses are the dominant source of error at low surface density,
below $\sim$20 M$_\odot$~pc$^{-2}$ or about $2 - 3 \times 10^{21}$ cm$^{-2}$
in column density.  As a result our observations do not offer much
insight into the physical nature of the scatter below those densities,
or for star formation surface densities below about 
0.01 M$_\odot$~yr$^{-1}$~kpc$^{-2}$; observations of other galaxies
in the future should reveal more about that surface density regime.  
However it is clear that at least some of the dispersion in the
Schmidt law above these scales is physical.  This is best seen at the
upper surface-density end of the plot, where the scatter clearly is
larger than the random observational errors, and is much larger
than the random uncertainties in the extinction-corrected
luminosities ($\pm$ 0.1 dex, see Fig.~3).  There are a number of
possible causes for this large scatter.  Variations in the ages
of the regions must be a factor; as a molecular complex evolves 
the ionizing flux will first peak then dissipate, and the cold
gas mass of the complex will evolve as well, as the region disperses
over time.  Moreover we have no reason to expect {\it a priori}
that the conversion fraction of gas to stars is a universal constant
in all clouds (see discussion in \S6).

None of the mechanisms discussed above are likely to bias the
slope of the SFR surface density vs gas surface density law to
a significant degree.  Another parameter that might influence
the dispersion or even the slope of the measured Schmidt law 
would be a large variation in the CO/H$_2$ conversion factor $X$.
A fixed value of $X$ has been adopted in this analysis.
The conversion factor would need to fluctuate by nearly an order of magnitude
to account for the observed scatter, and this is unlikely.
A multi-frequency study of M51a in CO by Garcia-Burillo 
et al. (1993) found evidence for possible variations in $X$ 
between the spiral arm and interarm regions, so we cannot rule
out some possible bias due to CO/H$_2$ variations.  However
we suspect that the dispersion mainly arises from a combination of
measuring uncertainties (especially in the molecular gas surface
densities) and physical effects including variations in 
the ages of the associations and clusters and actual
variations in the star formation efficiency among the clouds.

Figure 5 shows the correlation with the HI and inferred H$_2$
surface densities separately.  Molecular gas dominates most of
the gas clouds in the inner disk of M51a, so the comparison of
SFR and H$_2$ surface densities is similar to the relation
in Figure 4:

\begin{equation}
\log {\Sigma_{SFR}} = (1.37 \pm 0.03)~\log {\Sigma_{H_2}} - (3.78 \pm 0.09)
\end{equation}

\noindent
where the units for the SFR and hydrogen surface densities are the
same as in eq. (7).  The slope of this molecular-only relation is 
significantly shallower than for the SFR vs total (atomic $+$ molecular) 
surface density relation
($N = 1.37 \pm 0.03$ vs 1.56 $\pm$ 0.04); this arises because the atomic
gas contribution is proportionally larger in the lowest surface density
regions.

The strong correlation observed between the local SFR surface densities
and molecular gas surface densities in M51a is quite unlike
the relatively poor correlation between the disk-averaged SFRs and
molecular surface densities of normal spiral galaxies (e.g.,
Buat et al. 1989, Kennicutt 1989).  However our result is 
consistent with other spatially-resolved measurements 
of nearby galaxies, based on either point-by-point measurements
of azimuthally averaged radial profiles of SFR and gas surface
densities (Kennicutt 1989, Wong \& Blitz 2002, 
Heyer et al. 2004, Komugi et al. 2005).  These studies yielded
power law exponents $N$ between 1.3 and 1.4 when the SFR and
molecular gas surface densities are correlated.  Likewise 
Zhang et al. (2001) derive $N$ = 1.20--1.38 from an analysis
of star-forming regions in the Antennae (NGC~4038/9); their
fits apply to the total gas surface density, but since most
of the regions are dominated by molecular gas this is consistent
with the other results cited here.   Two other papers report
different results.  Kuno et al. (1995) carried out a point-by-point 
analysis of M51a using 16\arcsec\ beam CO observations with the Nobeyama
Radio Observatory along with published CO and H$\alpha$ data.
They derived a best-fitting Schmidt law slope $N = 0.7 \pm 0.1$.
The different result can be attributed the adoption of a
much lower CO/H$_2$ conversion factor in the Kuno et al. study
(1.0 $\times 10^{20}$ {\it vs} 
2.8 $\times 10^{20}$\,H$_2$\,({K\,km\,s$^{-1}$)}$^{-1}$ here), 
which is low enough for HI to be
the dominant component in many regions, and the absence of any  
extinction corrections in the (H$\alpha$) SFR measurements.
An analysis of radial profiles of 11 nearby spirals by
Boissier et al. (2003) derived significantly steeper ($N \sim 2$)
Schmidt law indices; the difference in this case can be
attributed to their use of a radially varying 
(metallicity-dependent) CO/H$_2$ conversion factor
(a constant factor was used in the other studies cited).
These comparisons underscore the dominant role of systematic
uncertainties such as the CO/H$_2$ conversion factor and
accurate extinction corrections in determining the form
of the observed SFR vs gas surface density relation in galaxies.

In contrast to the strong correlation seen in Figure 5 between
the SFR and molecular gas surface densities, there is virtually no
correlation between the local SFR surface density and the HI 
surface density.  We found this somewhat surprising, because
if the HI is formed by the photodissociation of molecular gas
by ambient stellar ultraviolet radiation (e.g., Shaya \& 
Federman 1987, Tilanus \& Allen 1991), one might expect the 
atomic surface density to scale with the local SFR density
(Allen et al. 1997).  In any case the lack of any clear
correlation between SFR surface density and HI surface 
density on local scales stands in stark contrast to the
relatively strong SFR vs HI correlation seen on global scales in disks
(e.g., Buat et al. 1989, Kennicutt 1989).  This difference probably
arises in part from the different molecular fractions in the two
cases.  Molecular gas comprises $>$90\%\ of the cold gas in M51a,
and is even more dominant in the center of the galaxy (Lord \&
Young 1990), so there HI is a trace species, especially in the
dense peaks where star formation takes place.  On the other hand
HI typically comprises $\sim$50\%\ of the cold gas in the disks
of the spirals studied by Kennicutt (1989, 1998b), and the objects
studied have a much larger range of SFRs and metallicities.   
Nevertheless the poor correlation between SFR and HI surface densities
in Figure 5 raises the interesting question of whether the global
correlation breaks down generally on subkiloparsec scales.  This
is a question we intend to pursue with studies of the larger SINGS sample.

Another interesting feature in Figure 5 is the presence of an
apparent upper limit to HI surface density, at about 
25~M$_\odot$~pc$^{-2}$, or a corresponding HI column density of
$\sim2 \times 10^{21}$~cm$^{-2}$.  Inspection of the HI map
shows that this is a general characteristic of the disk; a 
histogram of column densities shows a sharp falloff above this
value.  A similar behavior was seen by Wong \& Blitz (2002)
in an analysis of radial profiles of HI, CO, and H$\alpha$ for a subset of BIMA
SONG galaxies.  We suspect that this represents the column density above
which conditions in the clouds strongly favor the formation of
a dominant molecular medium.  Since most of the star formation 
in M51a takes place in denser molecular-dominated 
regions perhaps the lack of correlation
between SFR density and HI surface density should not be surprising.

The upper envelope of the SFR 
surface density vs gas surface density correlation tends to be dominated
by regions with relatively weak CO emission.
This is shown clearly in the right panel of Figure 4, 
where circles denote the positions of CO (3$\sigma$) upper limits.  
Many of these regions are also faint in HI, H$\alpha$, and the
infrared, and may be nothing more than small star forming
clouds that fall just below the detection limits of the BIMA
CO map.  Some of these could be evolved clouds, where 
star formation is well established and the parent molecular
clouds are dissipating.  This latter interpretation is supported
somewhat by the spatial distribution of the upper limit 
points.  As can be seen in Figure 2, the regions with CO
upper limits (blue circles) preferentially lie outside of
the main spiral arms.  However roughly a third of the points coincide
with or lie inside the main H$\alpha$ arms, so this evolutionary
hypothesis cannot be the sole explanation.  Otherwise we
did not detect any systematic dependence of the Schmidt law
zeropoint on arm position, but this is hardly surprising in
view of the observational uncertainties in the gas surface
densities.

Finally, Figure 6 shows the same data as plotted in Figure 4,
but here with the points coded in color by 
galactocentric radius.  This is useful for checking whether the
Schmidt law itself could be dependent on radius, and also whether
there are any hints of other radially dependent systematic effects
in the data.  The comparison shows that the Schmidt laws at different
radii largely overlap with each other; there is no evidence for any
significant radial dependence.  The only possible exception is the
strong clustering of points at high SFR and gas density at the
very smallest radii (0.5--2 kpc),
where there is a hint of a turnover in the power law.  
This could arise from a number of measurement effects, such
as a change in the CO/H$_2$ conversion factor at the highest metallicities
or a significant absorption of ionizing photons by dust in the 
dustiest central regions, or from a physical effect,
as introduced for example by a change in disk kinematics 
in the central regions.

\subsection{The Star Formation Law on Other Linear Scales}

It is interesting to examine whether the form of the star formation
law changes significantly as a function of the physical scale over
which the SFRs and gas densities are correlated.  Here we combine
our data with other single-dish studies to explore such variations
on scales of $\sim$0.3 $-$ 1.8 kpc.

As discussed earlier the resolution of our data prevent us from 
reliably probing the form of the star formation
law on scales less than 300 pc in M51a.  As an exploratory
exercise we carried out a set of measurements using aperture
diameters of 7\farcs3 (300 pc), the smallest aperture for which we
felt we could reliably measure fluxes, given the beam sizes of our
24~\micron, HI, and CO measurements.
As before we centered the apertures on the emission peaks in order to obtain
reliable photometry.  The resulting SFR and gas densities tend to
shift to higher values (because the surface densities are more centrally
concentrated), but the distribution of
points closely follows that of the 13\arcsec\ data, with a somewhat
larger dispersion about the mean relation.
This suggests that any transition in the form of the star
formation law from a nonlinear power law relation must occur
on scales considerably smaller than 300 pc.  However we are 
reluctant to attach much physical significance to this result,
because the apertures are at the limit of the resolution of our 
infrared, CO, and HI data, and sensitivity limits at this resolution
forced us to measure only the brighter star formation peaks, mainly
in the spiral arms.  The consistency of results is
interesting and needs to be followed up on more nearby galaxies
where higher spatial resolution can be achieved.

We have also used the FCRAO single-dish CO data of Lord \&
Young (1990) to examine the star formation law with aperture
diameters of 45\arcsec\ (1850 pc).   They obtained measurements
at 60 positions, and these cover virtually all of the disk out
to the edge of the main spiral pattern ($\sim$5\arcmin\ diameter).
We used the CO measurements from their paper, and also measured
CO fluxes from the BIMA SONG maps using their aperture positions
and sizes, for 58 objects in common between the two map sets.
The two CO datasets give consistent results, but we give
preference to the FCRAO data because they have higher
signal/noise on these extended scales.
We applied the same apertures to our data to measure corresponding
HI, H$\alpha$, and 24~\micron\ fluxes.  

The result of this comparison is shown in Figure 7, which again
shows the relationship between SFR surface density and (total)
hydrogen surface density, but measured in this case with 1850 pc
diameter apertures that fully sample the disk of M51a.
Again a strong correlation is observed, with a best-fitting relation:
$\log {\Sigma_{SFR}} = (1.37 \pm 0.03) \log {\Sigma_H} - (3.90 \pm 0.07) $.
The slope of this relation matches within  
the uncertainties the value of $N = 1.4 \pm 0.15$ 
seen in global measurements of galaxies (Kennicutt 1998b, \S5.3).
However the slope of the relation is somewhat shallower 
than that measured in the 520 pc aperture data (where 
$N = 1.56 \pm 0.04$), and the zeropoint of the relation is 
significantly higher, by approximately 0.4 dex (see eq. [7] and Fig.\ 4).
As discussed below (\S5.3) these differences in relations can 
be attributed to different beam filling factors in the respective
sets of measurements.  Despite these differences it is clear that
a Schmidt power law provides a good parametrization of the SFR
on scales extending from 300 $-$ 1850 pc, out to integrated measurements
of disks.

The scatter in the 1850 pc relation is roughly a factor of two
lower than the corresponding Schmidt law on 520 pc scales.
Presumably this results from the averaging
over large numbers of individual regions in the larger-aperture
measurements.  The scatter about the best-fitting relation in
Figure 7  ($\pm$0.24 dex rms) is larger than the estimated
random error in the gas and SFR densities, but it is not 
significantly larger when systematic errors in the measurements
are taken into account, especially when including the extinction
corrections.

\subsection{Comparison with the Global Schmidt Law for Galaxies}

This study was motivated by the discovery of a surprisingly
strong and tight Schmidt law relating the disk-averaged SFR surface densities
and gas surface densities of galaxies, extending from normal spirals to 
luminous infrared starburst galaxies (Kennicutt 1998b).  So an
obvious question is how the local law we have measured in M51a
compares to this global relation between entire galaxies.
The comparison is shown in Figure 8.  Plotted are the SFR and
gas surface densities for the 520 pc and 1850 pc data (open circles
and solid triangles, respectively).  The best fitting solution for
the 520 pc data is shown as the solid line, while the dashed line
shows the corresponding fit to the 1850 pc data.  
The dotted line shows the best fit to the integrated Schmidt law
for normal galaxies and infrared-selected
starburst galaxies from Kennicutt (1998b).  Finally the solid blue
square in Figure 8 shows the mean integrated SFR and gas surface 
densities for M51a from the Kennicutt (1998b) analysis.

As expected the local relations in M51a are qualitatively consistent
with the global law, but there is a significant offset, with the
M51a relations lying lower by 0.46 and 0.39 dex, for the 13\arcsec\ and
45\arcsec\ measurements, respectively.  We need to bear in mind
that the SFR and gas surface densities measured for individual
subregions cannot be defined in a way that is entirely consistent with
disk-averaged measurements of galaxies; the sample is biased to
actively star-forming regions and gas density peaks, the SFR 
calibrations are different, and the surface area used to convert
from SFRs to SFR surface density is somewhat arbitrarily selected.
So we would be startled if the relations corresponded exactly,
but despite that we find the offset of $\sim$0.4 dex to be surprising.
In the Kennicutt (1998b) study M51a lies 0.24 dex below the overall 
galaxy sample fit, which may account for part of the difference.

Most of the remaining difference can be attributed to the filling
factor of star-forming regions in the disk of M51a.  As an illustration,
consider an idealized case of a disk containing $n$ identical
star-forming regions, each with size $r$, star formation
rate $\psi$, and gas mass $M_g$.  The disk itself has a radius $R_d$.  
The SFR densities and gas densities of the star-forming regions themselves 
are simply $\psi/\pi r^2$ and $M_g/\pi r^2$, whereas the corresponding
SFR and gas surface densities
averaged over the entire disk are $n\psi/\pi R{_{d}^{2}}$ 
and $n {M_g}/\pi R{_{d}^{2}}$, 
respectively.  Both densities are offset to lower values by the
same factor $n {r^2}/R{_{d}^{2}}$.  However because the slope of the Schmidt
law is steeper than a linear relation, the effect of the larger
beam sampling will be to offset the disk-averaged densities away
from the spatially-resolved Schmidt law.  For the case of a Schmidt
law with slope $N \sim 1.5$, the approximate offset will be 
the square root of the individual surface density offsets, or
a factor ${n^{0.5}} r/{R_d}$.  In the case of M51a, $n = 257$, $r$ = 6\farcs5, 
and $R_d$ = 300\arcsec, and thus we predict that the global relation should
be offset from the spatially-resolved relation by $\sim$0.46 dex.
This idealized calculation actually overestimates the offset because
in reality there is a considerable amount of star formation and
gas at low surface brightness located outside of the 257 regions
we measured.  When all factors are taken into consideration the
observed offset is approximately in agreement with what we would
expect from the aperture bias.  This same effect can account
for the slight offset in zeropoint between the Schmidt law fits
to the 520 pc and 1850 pc apertures (\S5.2), because the latter measurements
cover the inner disk of M51a, so on average the beam filling factor
derived for the entire disk applies.   

There also is a significant difference in slopes between the 3
relations that are plotted in Figure 8, ranging from
$N = 1.56 \pm 0.04$ for the 520 pc M51a measurements to
$N = 1.37 \pm 0.03$ for the 1850 pc M51a data and
$N = 1.40 \pm 0.15$ for the global galaxy law.  The uncertainties
quoted for the M51a measurements only include random errors,
while the uncertainty given for the global law is dominated
by systematic errors, mainly possible systematic variation
in the CO/H$_2$ $X$-factor over the large range in gas densities
and radiation field environments over which the global law applies.
For example a change in $X$ by a factor of two between the 
IR-luminous starburst galaxies and normal galaxies would be sufficient
to increase the slope of the best-fitting global law from $N = 1.4$
to 1.5 (see Kennicutt 1998b).  As discussed earlier similar
effects may introduce systematic shifts into the relations
derived for M51a.  In addition, the
aperture sampling effects discussed above can also introduce
a second-order change in the slope of the Schmidt law if the
filling factor of HII regions changes systematically as a 
function of SFR and gas surface density.  For example the 
fraction of the 45\arcsec\ beams containing star-forming
regions varies from about 10--100\%\ in M51a, with most of the
sparsely populated positions occurring at the lowest gas and
SFR surface densities.  This can shift the slope of the 1850 pc
aperture relation by up to $-$0.2 dex, consistent with the slope
offset we observe.  As a result, when one takes into account
these possible systematic errors the actual uncertainties in
the Schmidt law slopes derived for M51a are at least $\pm$0.1
in $N$, and hence we are reluctant to 
attach any astrophysical significance to the differences between
the relations seen in Figure 8, until we have an opportunity to
study the local relations in more galaxies and construct an
improved global relation.

\subsection{Alternate Forms of the Star Formation Law}

As discussed in \S1 the global SFR and gas surface densities
of galaxies can be fitted to relations other than a Schmidt law,
including the scaling with gas density divided by mean dynamical
time (eq. [2]).  How well do the resolved observations of M51
fit such a relation?  We show the comparison in Figure 9, which
plots the SFR densities as a function of the ratio of hydrogen
density (HI $+$ H$_2$) to orbit time for that cloud 
($2 \pi R/V_{rot}$).  We have used different colors to denote 
the 4 ranges in galactocentric radii, as in Figure 6.
The SFR densities and gas densities were calculated as described
earlier, and the orbit times were computed using the M51 rotation
curve from Sofue et al. (1999).  For this model to be valid the slope of
the relation is constrained to be unity, so the solid line
shows the unit slope line that bisects the data points.  Also
shown as the dashed line is the global dynamical time relation
from Kennicutt (1998b).  

Figure 9 reveals a general, qualitative trend for the regions with highest
ratio of density to orbit time to have higher SFRs.  There is
a strong radial segregation of points in this plot, due to the
roughly $1/R$ falloff in orbit time over most of the disk.
However the slope of the mean relation is far from unity 
($\sim$0.65), and the scatter about the mean relation is very 
large ($\pm$0.4 dex), though not significantly higher than the scatter in
the Schmidt law discussed earlier.  This relation is also
offset below the comparable global relation in Kennicutt (1998b)
in this case by a factor of 5 (0.7 dex).  We tentatively conclude
that although this kinematic star formation law may have some
usefulness for characterizing the integrated star formation in
galaxies and starbursts, it may be less useful as a description
of local star formation in galaxies.
We intend to explore this much more carefully when results from
the full SINGS dataset are analyzed.  

\subsection{Evidence for Star Formation Thresholds?}

Previous spatially-resolved observations of star formation in
galaxies have provided a large body of evidence suggesting that
the monotonic behavior of the star formation law at high gas surface 
densities shows a break at low surface densities, usually characterized
as a star formation threshold (e.g., Kennicutt 1989, 1997; Martin
\& Kennicutt 2001, and references therein).  These thresholds
have been ascribed to a variety of physical mechanisms, including
large-scale gravitational instabilities (e.g., Quirk \& Tinsley 1973;
Zasov \& Simakov 1988; Kennicutt 1989; Hunter et al. 1998; Elmegreen 2002),  
or molecular or cold gas phase formation thresholds (e.g., Elmegreen
\& Parravano 1994; Schaye 2004; Blitz \& Rosolowsky 2004).
Our spatially resolved data allow us to check for the observational
signatures of thresholds.  In particular we can compare the local gas
surface densities with the predicted threshold densities for gravitational
instability, and test whether the observations are consistent with that
picture.   

Examination of Figures 4--7 shows little evidence for any star formation
thresholds.  The only possible hint might be a handful of regions
with the lowest observed SFR surface densities ($\log \Sigma_{SFR} < -2.6$);
most of these points lie well below the extrapolated Schmidt law fit,
as would be expected if they lay below a threshold.  However we believe
that most of that trend is due to the sensitivity limit of the CO maps.
If there are regions of the disk with lower SFR and 
gas surface densities they would not be detected in our data.

To test further for threshold effects we calculated for each region 
the expected critical density using the relation of Kennicutt (1989),
which is based on applying the Toomre (1964) gas stability criterion
for an isothermal disk of gas clouds:

\begin{equation}
\Sigma_c = { \alpha {{\kappa c} \over {\pi G}} }
\end{equation}

\noindent
where $\Sigma_c$ is the critical (total) gas surface density for 
star formation, $\kappa$ is the epicyclic frequency, $c$ is the velocity
dispersion of the gas (taken as 6 km~s$^{-1}$ following Kennicutt 1989),
and $\alpha$ is a scaling constant fitted to the observations 
(taken as 0.7 following the same paper), in order to reproduce the
observed H$\alpha$ edges of nearby galaxies.  The values of $\kappa$ were
calculated from the rotation curve of Sofue et al. (1999).
Figure 10 shows the distribution of these threshold normalized
surface densities, which correspond roughly to $1/Q$ in terms
of the Toomre stability index $Q$. It is interesting that the distribution 
shows a strong turnover below a value of unity ($Q > 1$), 
where one would expect if our sample
is limited to regions with active star formation.  This result
is hardly robust enough to provide firm evidence for thresholds,
but its general consistency with the $Q$-threshold picture is interesting.
Of the 257 regions, 29 show local gas surface densities that are
below the expected threshold, yet they are forming stars.  It is
possible that we are seeing a breakdown of the simple threshold
model in these cases, but unfortunately they each deviate by less
than 1$\sigma$ of $\Sigma_c$; given the large number of points near
$\Sigma_c$ we may well be observing nothing more than the spillover
of observational errors in the tails of the distribution.  In short
our data do not extend deep enough to offer a concise test of the
gravitational threshold model, and all that we can say is that
the observations are roughly consistent with expectations from that model.

The distribution of HI surface densities elsewhere in M51a (where star
formation is not observed) lies almost entirely below the 
$\Sigma_{gas}/\Sigma_c = 1$ limit, again consistent with the gravitational
thresold picture.  However we are reluctant to attach much significance
to this result, because over much of the disk the CO sensitivity
limit lies close to the expected threshold density, so it is difficult
to disentangle this incompleteness from a threshold effect.  
We expect to be able to make more critical tests for threshold effects
in some of the other galaxies in the SINGS/SONG sample.
In particular, a comparison of the star formation law for the 
spiral arm and interarm regions is being carried out by
de Mello et al. 2007 (in preparation).  

\section{Discussion and Summary}

Our main result is that on spatial
scales extending down to at least 500 pc, the SFR surface density
is correlated, at least in a statistical sense, with the local
gas surface density, following a Schmidt power law:

\begin{equation}
\log {\Sigma_{SFR}} = (1.56 \pm 0.04)~\log {\Sigma_H} - (4.32 \pm 0.09)
\end{equation}

\noindent
where $\Sigma_{SFR}$ is measured in units of 
M$_\odot$~yr$^{-1}$~kpc$^{-2}$ and $\Sigma_H$ is measured in units
of M$_\odot$~pc$^{-2}$, and these quantities are sampled with
circular apertures of 520 pc diameter.  This equation was fitted to the total
(atomic plus molecular) hydrogen surface densities, but in M51a the 
correlation with molecular surface density alone is very similar, with
$\log {\Sigma_{SFR}} = (1.37 \pm 0.03)~\log {\Sigma_{H_2}} - (3.78 \pm 0.09)$,
as detailed in \S5.1.  The uncertainties quoted only include formal
fitting errors, and do not incorporate any possible systematic
errors.  If we consider the 1850 pc aperture measurements as
a largely independent measurement of the Schmidt law in M51a,
then we can use the difference in fits of the 520 pc and 1850 pc
data as providing a more realistic indication of the uncertainties.

%

As we stated at the outset of the paper, this study was mainly 
intended to build the methodological foundation for a larger study
of the SINGS sample in future papers.  A key element was the
calibration of a combined infrared plus H$\alpha$ star formation
index, which provides more precise H$\alpha$ extinction
corrections for HII regions than have generally been available
previously.  This in turn makes it possible to 
quantify the form of the SFR vs gas surface density law on a point-by-point
basis in galaxies.  We are also extending this basic approach of
multi-wavelength SFR tracers to more luminous starbursts 
(Calzetti et al. 2007) and to galaxies as a whole (Kennicutt et al.
2007, in preparation).  

This analyis has revealed other interesting results.  We find 
that the same type of power law relation that describes the global
SFRs of galaxies appears to reproduce the star formation law
down to local scales of 300 $-$ 1850 pc, approaching at the low extremes
the scales of individual giant molecular cloud complexes.
Although these relations are defined in terms of total (atomic plus
molecular) gas surface densities, in M51a they mainly trace an underlying
correlation with the molecular surface density component.
This must result at least partly from the dominance of molecular gas in M51a.
By contrast the local SFR surface density is virtually uncorrelated
with the surface density of HI on these scales.  

As mentioned earlier, when the global, disk-averaged SFR surface densities
of galaxies are correlated with the disk-averaged atomic, molecular,
and surface densities, the strongest correlations are with the total
gas density (Buat et al.\, 1989, Kennicutt 1989, 1998b).  Indeed
among normal star-forming disk galaxies the SFR surface density is 
only weakly correlated at best with the CO-inferred molecular surface
density (Kennicutt 1989), quite the opposite of what is observed here on
a point-by-point basis in M51a.  On the other hand, in infrared-luminous
starburst galaxies, which typically contain dense compact gas disks,
the SFR and molecular surface densities are tightly correlated.
So it may well be that the behaviors of the spatially-resolved and
disk-integrated SFR vs molecular surface density relations are consistent
when similar regimes in surface density are compared.  This raises
a separate question of whether the tightness of the 
$\Sigma_{SFR}$ {\it vs} $\Sigma_{H_2}$ at high surface density arises
because of a fundamental correlation between the SFR and the 
molecular gas phase, or alternatively from an underlying correlation 
of the SFR with the total gas density, which manifests itself as
a correlation with $\Sigma_{H_2}$ only when the gas is predominantly
molecular?  We hope to address this question by extending our analysis
to galaxies with a larger atomic gas component.

Although we have observed broad consistency in the form of the
Schmidt law across a wide range of physical scales, as discussed
in \S5.3 we do observe significant shifts in the zeropoint 
(and possibly the slope) with scale size.  Our results
suggest that the correlation between SFR and gas surface densities
on small scales defines an intrinsic Schmidt law, and when
these surface densities are measured with larger measuring apertures
(which include an increasing fraction of area devoid of star-forming
regions and gas), the zeropoint of the Schmidt law becomes larger, 
because of the nonlinear slope of the relation.   
So which relation is more fundamental?
It is tempting to define the spatially-resolved relation as
the physically fundamental one, but this relation is based on
a highly biased subsampling of the disk, limited to the most
massive GMC complexes and giant HII regions.  The larger aperture
measurements provide a completely unbiased sampling of the disk,
but are based on averages of SFR and gas surface densities which vary
locally by orders of magnitude within the measuring apertures.  The
most important lesson is that this scale dependence of the Schmidt
law must be taken into account when it is applied to a dataset
or to a theoretical model.  For example, our results show that
the SFR surface density predicted for a region of fixed gas surface
density can differ by more than a factor of two, depending on whether
the size of the region of interest is $\sim$0.5 kpc or averaged
over the entire disk of a galaxy.

Eventually we hope that data of this kind will shed new insights into
the physical origins of the observed star formation law.
A rigorous {\it ab initio} theory for star formation on these scales
is not yet in place, so it is not entirely clear what theory would
predict for the form of the local star formation law.   Nevertheless
our observations provide some tantalizing clues.  
Since our measurements have been made with fixed-diameter apertures,
the gas surface densities can be readily converted to total
gas masses, and the combination of H$\alpha$ and infrared
luminosities provides a direct measurement of the ionizing flux
of the embedded stars.  The nonlinearity in the observed Schmidt
law thus implies that the present instantaneous SFR per unit
gas mass increases in the more massive clouds (or complexes).
It is very tempting to attribute this result to a possible
increase in the star formation efficiency in more massive
clouds, that is, a higher fraction of stars formed in the
more massive clouds.  However this direct extrapolation 
is not valid, because the measured ionizing fluxes only 
provide information at most on the mass of recently formed
O-stars in the clouds, and not on the total mass of stars
(of all stellar masses) formed over the lifetimes of the 
clouds.  One could explain a $N \sim 1.5$ Schmidt law even
if the star formation efficiency were the same for all
cloud masses and gas surface densities, if for example
the star-forming lifetimes of massive clouds were 
systematically lower than for low-mass clouds, or if 
the period of peak formation of O-stars decreased with
increasing cloud mass.  Without further observational
constraints on these time scales one cannot draw any
direct association between the slope of the star formation
law and the constancy (or not) of the cloud-averaged 
star formation efficiency.  However the extension of the
nonlinear Schmidt law down to linear scales of 500~pc and 
cloud mass scales of order 10$^6 - 10^7$ M$_\odot$ 
strongly hints at either an increasing
star formation efficiency or a shorter star formation 
time scale with increasing cloud mass.  An important next
step would be an extension of this analysis to nearer 
galaxies with lower limiting cloud masses and SFRs, and 
to Galactic clouds, where direct information on stellar 
ages is available.

This case study of M51a has illustrated the value of
spatially-resolved infrared, H$\alpha$, HI, and CO observations of 
nearby galaxies for constraining the form and physical nature
of the star formation law.  
However, much future work is needed on this problem.  Within the larger
SINGS project we plan to extend this analysis to approximately 
15 other galaxies for which high quality CO, HI, H$\alpha$, and
Spitzer 24~\micron\ data are available.  These galaxies cover a
wide range of types and gas disk properties, and the extended
physical coverage may resolve some of the questions and selection
effects that have muddied the interpretation of these data.
Looking further ahead, the study of the star formation law in
galaxies remains limited in large part by the spatial resolution
and sensitivity of the molecular gas data, even with
the superb BIMA data in hand.  Follow-up deeper mapping
of a handful of galaxies, extending to limiting column densities
below those expected for gravitational stability would allow for
a much more physically meaningful interpretation of the observed
SFR law.  Finally, independent measures of extinction in some of
these galaxies (redundant with the infrared + H$\alpha$ extinctions
derived here) would provide much better constraints on the 
random and systematic measurements in our SFR measurements and
the dispersion of the star formation law.  The results of such
efforts will have far-reaching applicability to the understanding
of star formation in galaxies and the formation and evolution of
galaxies.

\acknowledgements

We would like to acknowledge valuable discussions with a number of
colleagues, including Ayesha Begum, Hsiao-Wen Chen,
Cathie Clarke, Crystal Martin, Phil Solomon, and Art Wolfe.
We thank the anonymous referee for a careful reading of the
manuscript.  
This work is based in part on observations made with the Spitzer Space 
Telescope, which is operated by the Jet Propulsion Laboratory, California 
Institute of Technology under a contract with NASA. Support for this work 
was provided by NASA through an award issued by JPL/Caltech.


\clearpage

\begin{deluxetable}{rcccccccccc}
\tabletypesize{\footnotesize}
\tablewidth{0pc}
\rotate
\tablecaption{Aperture Photometry\tablenotemark{1}}
\tablehead{
\colhead{ID}  & 
\colhead{RA} &
\colhead{Dec} & 
\colhead{D$_{ap}$} &
\colhead{{$\log$\,L$_{H\alpha}$}\tablenotemark{2}} & 
\colhead{{$\log$\,L$_{Pa\alpha}$}\tablenotemark{2}} &
\colhead{{$\log$\,L$_{24}$}\tablenotemark{3}} & 
\colhead{{$\log$\,$\Sigma_{SFR}$}\tablenotemark{4}} & 
\colhead{$\log$\,N$_{HI}$} & 
\colhead{$\log$\,N$_{H2}$} & 
\colhead{$\log$\,$\Sigma_H$} \\
\colhead{} & 
\colhead{J2000} & 
\colhead{J2000} & 
\colhead{\arcsec} & 
\colhead{ergs\,s$^{-1}$} & 
\colhead{ergs\,s$^{-1}$} & 
\colhead{ergs\,s$^{-1}$} & 
\colhead{} & 
\colhead{H\,cm$^{-2}$} &
\colhead{H$_2$\,cm$^{-2}$} & 
\colhead{M$_\odot$\,pc$^{-2}$} \\
}
\startdata
   1 & 13:29:51.5 & 47:11:43 & 13 & 38.86 & 38.84 & 41.17 & -0.50$\pm$0.025 & 20.26 & 21.98 & 2.19$\pm$0.094 \\
   2 & 13:29:52.8 & 47:11:33 & 13 & 39.09 & 38.83 & 41.16 & -0.57$\pm$0.026 & 19.65 & 22.00 & 2.21$\pm$0.090 \\
   3 & 13:29:52.3 & 47:11:27 & 13 & 38.90 & 38.78 & 41.09 & -0.57$\pm$0.028 & 20.68 & 22.17 & 2.38$\pm$0.063 \\
   4 & 13:29:51.8 & 47:11:23 & 13 & 38.74 & 38.67 & 40.92 & -0.68$\pm$0.037 & 20.65 & 22.10 & 2.31$\pm$0.073 \\
   5 & 13:29:53.1 & 47:11:21 & 13 & 38.73 & 38.44 & 40.99 & -0.96$\pm$0.061 & 20.76 & 22.23 & 2.44$\pm$0.055 \\
   6 & 13:29:53.8 & 47:11:24 & 13 & 38.86 & 38.59 & 41.08 & -0.80$\pm$0.044 & 20.46 & 22.28 & 2.49$\pm$0.050 \\
   7 & 13:29:52.7 & 47:12:03 & 13 & 39.04 & 38.92 & 41.24 & -0.45$\pm$0.021 & 20.71 & 22.36 & 2.57$\pm$0.042 \\
   8 & 13:29:52.2 & 47:11:58 & 13 & 39.19 & 39.04 & 41.36 & -0.33$\pm$0.016 & 20.62 & 22.47 & 2.68$\pm$0.032 \\
   9 & 13:29:52.0 & 47:12:05 & 13 & 39.17 & 38.95 & 41.32 & -0.43$\pm$0.019 & 20.64 & 22.31 & 2.52$\pm$0.046 \\
  10 & 13:29:51.4 & 47:12:02 & 13 & 39.11 & 38.91 & 41.36 & -0.47$\pm$0.021 & 20.77 & 22.39 & 2.60$\pm$0.039 \\
\enddata
\tablenotetext{1}{A complete listing of data in this table, including 13\arcsec\
          and 45\arcsec\ aperture photometry can be found in the on-line edition.} 
\tablenotetext{2}{Ionized gas luminosities have been corrected for Galactic
          foreground extinction ${E(B-V)}_{MW} = 0.037$, but otherwise have
          not been corrected for extinction.} 
\tablenotetext{3}{Defined as $\nu L_\nu$ at 24\,$\mu$m}
\tablenotetext{4}{In units of M$_\odot$\,yr$^{-1}$\,kpc$^{-2}$.
          SFRs for objects with Pa$\alpha$ photometry 
          were corrected for extinction using the Pa$\alpha$/H$\alpha$ ratio.
          Otherwise SFRs were corrected for extinction using the 24\,$\mu$m/H$\alpha$ 
          ratio, as described in the text.} 
\end{deluxetable}

\clearpage

\begin{figure}
\plotone{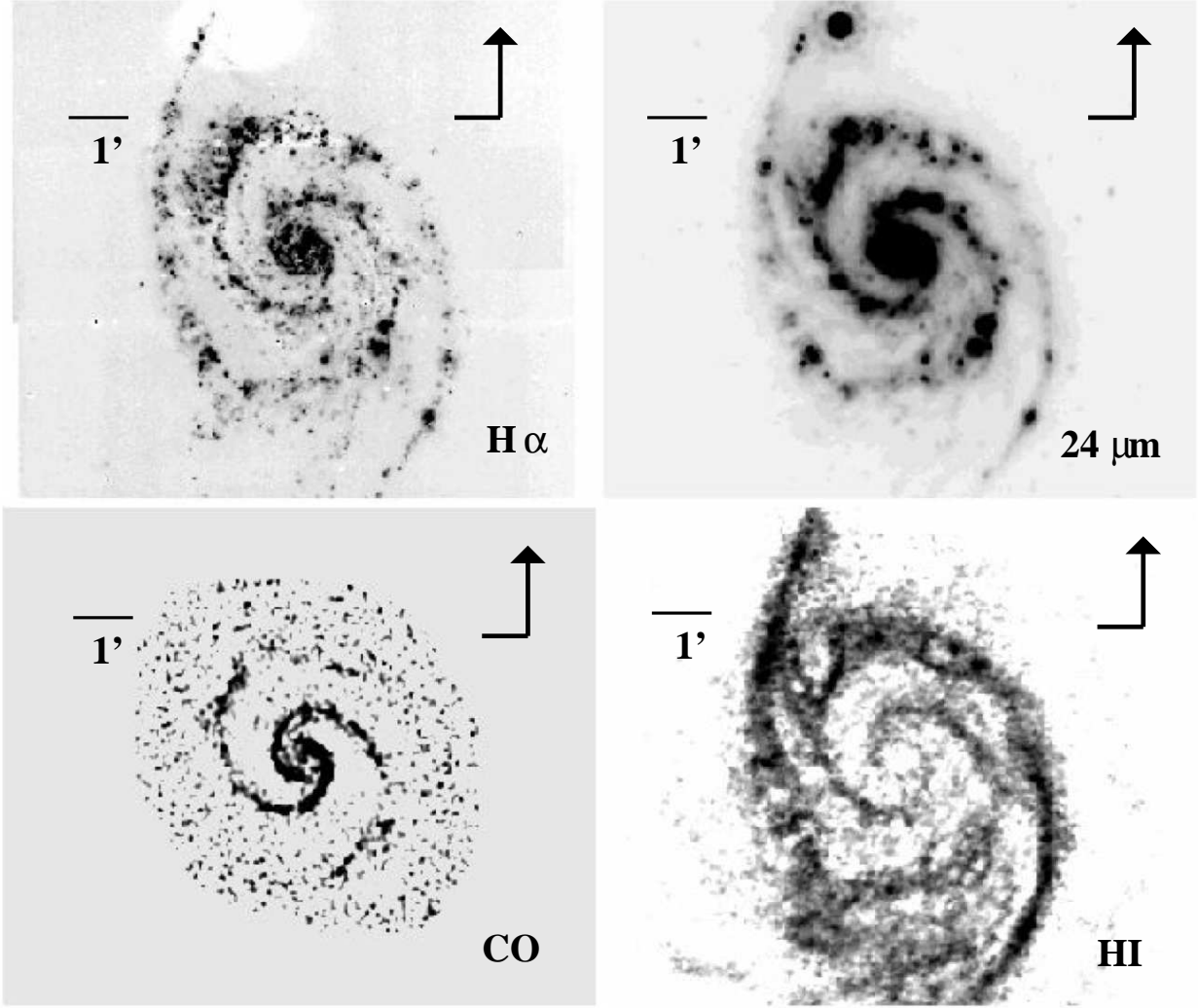}
\caption{M51 as observed in H$\alpha$, 24~\micron\ continuum, 
CO, and HI.  North is up (arrow) and east is to the left.
The horizontal bars in the left corner of each panel indicate an angular
scale of 1 arcminute (2.5 kpc).  These are shown to illustrate qualitatively
the relative distributions of the cold gas and star formation tracers.}
\end{figure}

\clearpage

\begin{figure}
\plotone{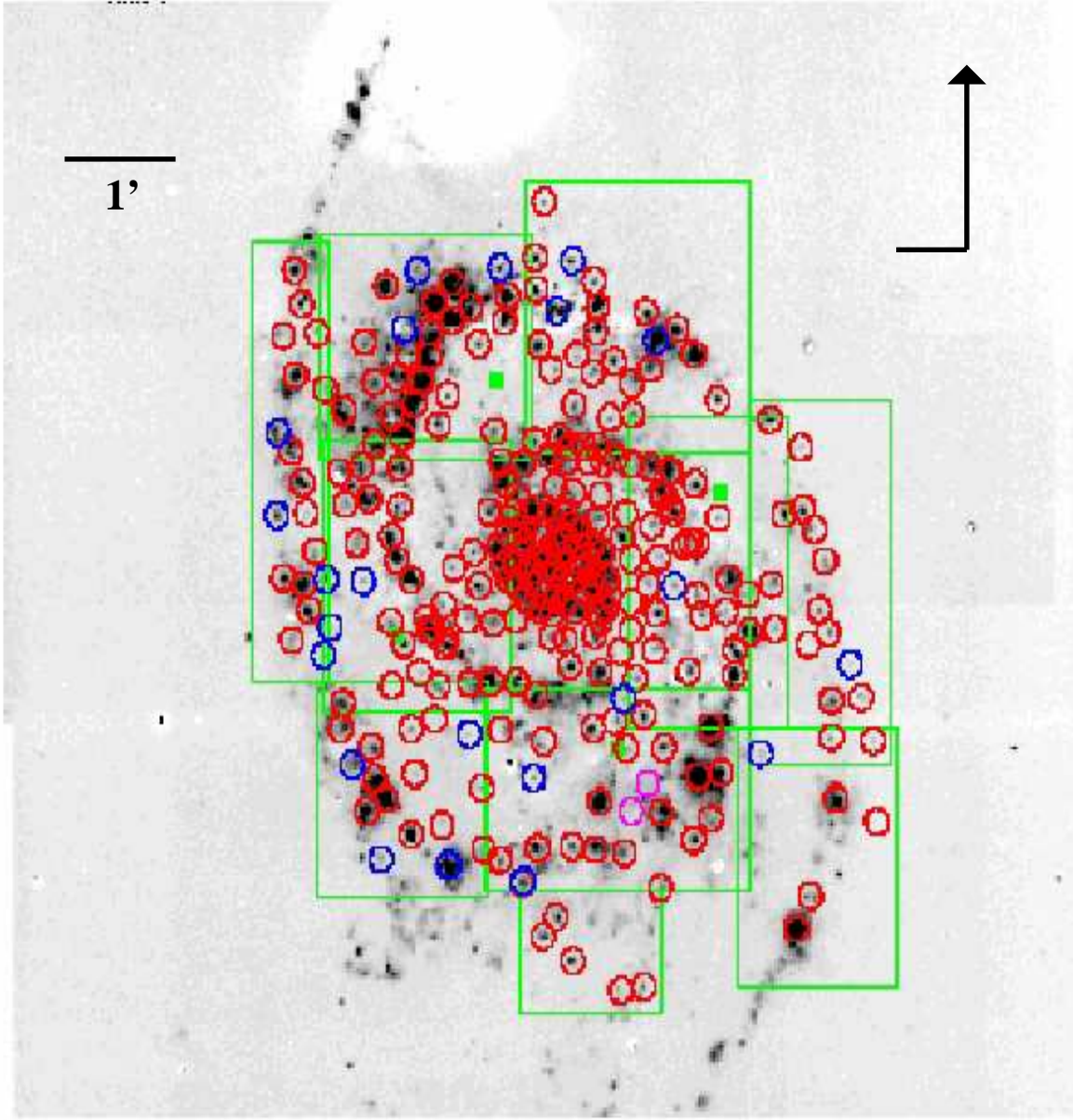}
\caption{Continuum-subtracted H$\alpha$ image of M51 with the 
13\arcsec\ (520 pc) measuring apertures shown.  
Red circles denote regions detected in CO, HI, H$\alpha$, and 24~$\mu$m.
Blue circles denote regions with CO upper limits (but detected at all
other wavelengths), and the two magenta circles denote CO sources
without significant detections of H$\alpha$ or infrared emission.
The rectangles show regions selected
for background determinations, as described in the text.}
\end{figure}

\clearpage

\begin{figure}
\epsscale{1.1}
\plottwo{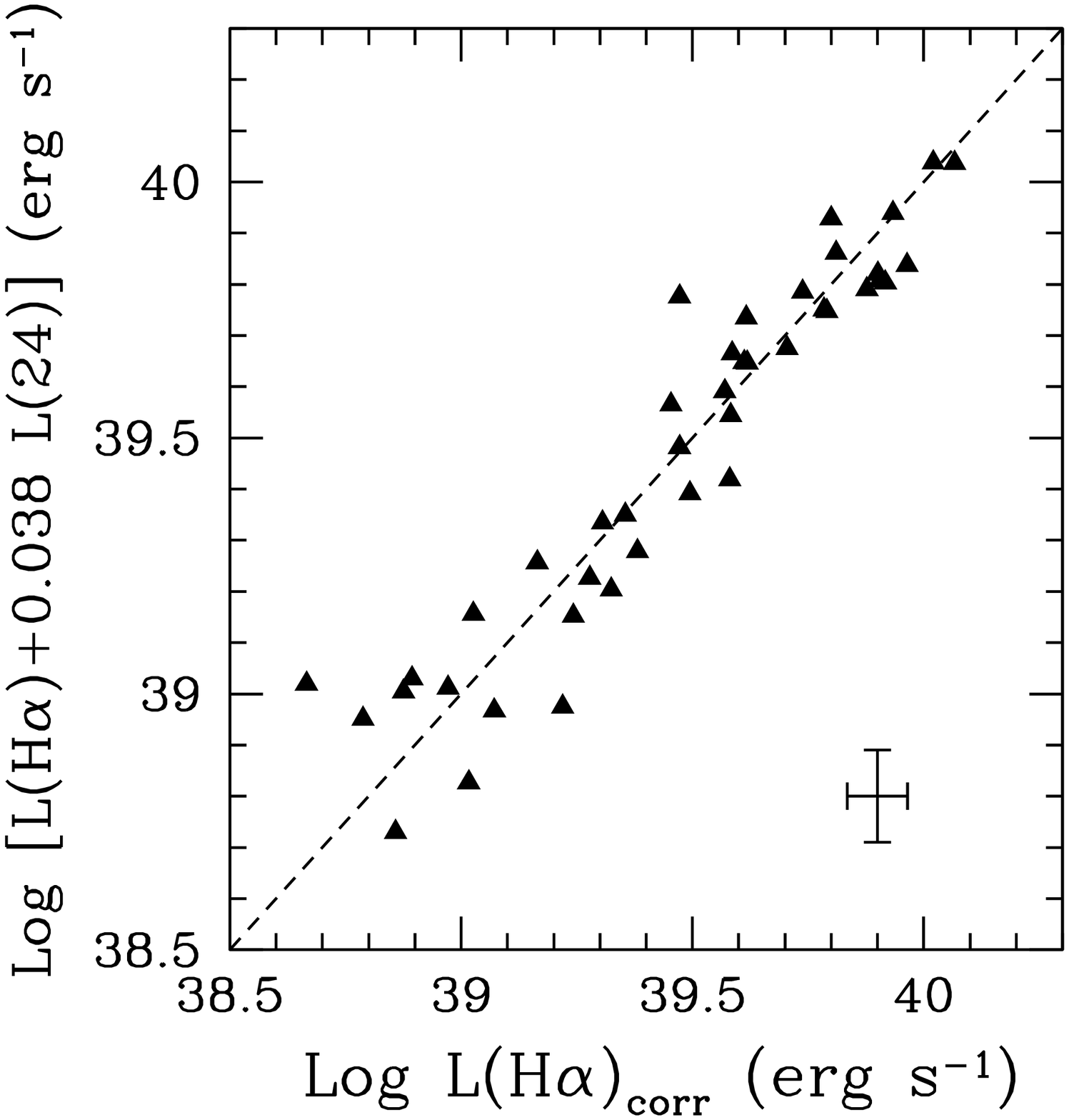}{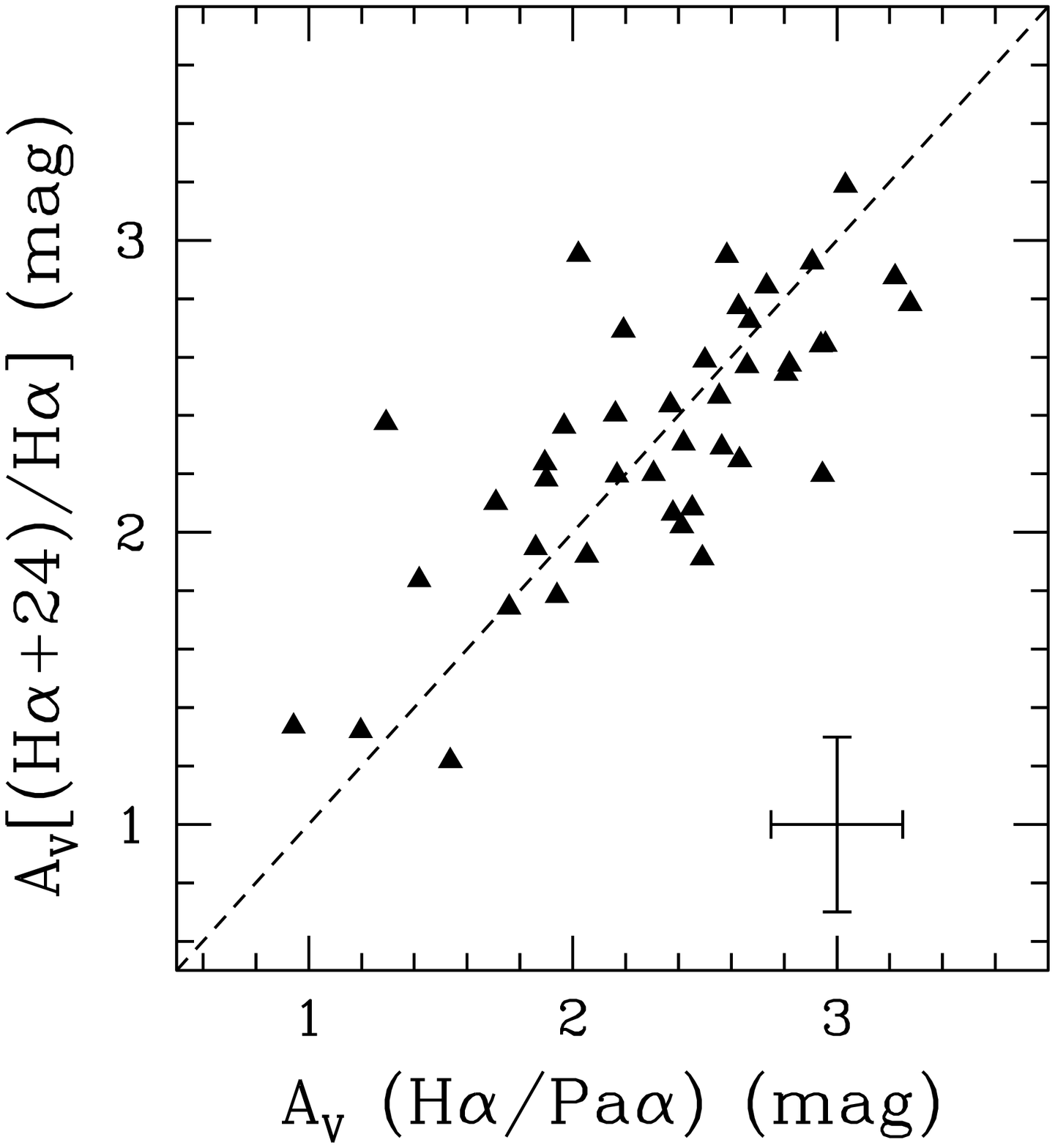}
\caption{Left:  Comparison of attenuation-corrected H$\alpha$ luminosities for
42 HII regions in the inner disk of M51, using a weighted sum of 
observed H$\alpha$ and 24~\micron\ luminosities, vs independently  
extinction-corrected H$\alpha$ luminosities, derived from the 
H$\alpha$/Pa$\alpha$ flux ratios.  The line shows the
median fit with a slope forced to unity.  The error bars show
typical uncertainties for the measurements.  Right:  Comparison of the
corresponding visual extinctions for the same objects, as derived from the
weighted sum of H$\alpha$ and 24~\micron\ fluxes, vs those derived
from H$\alpha$/Pa$\alpha$ flux ratios.}
\end{figure}

\clearpage

\begin{figure}
\epsscale{1.1}
\plottwo{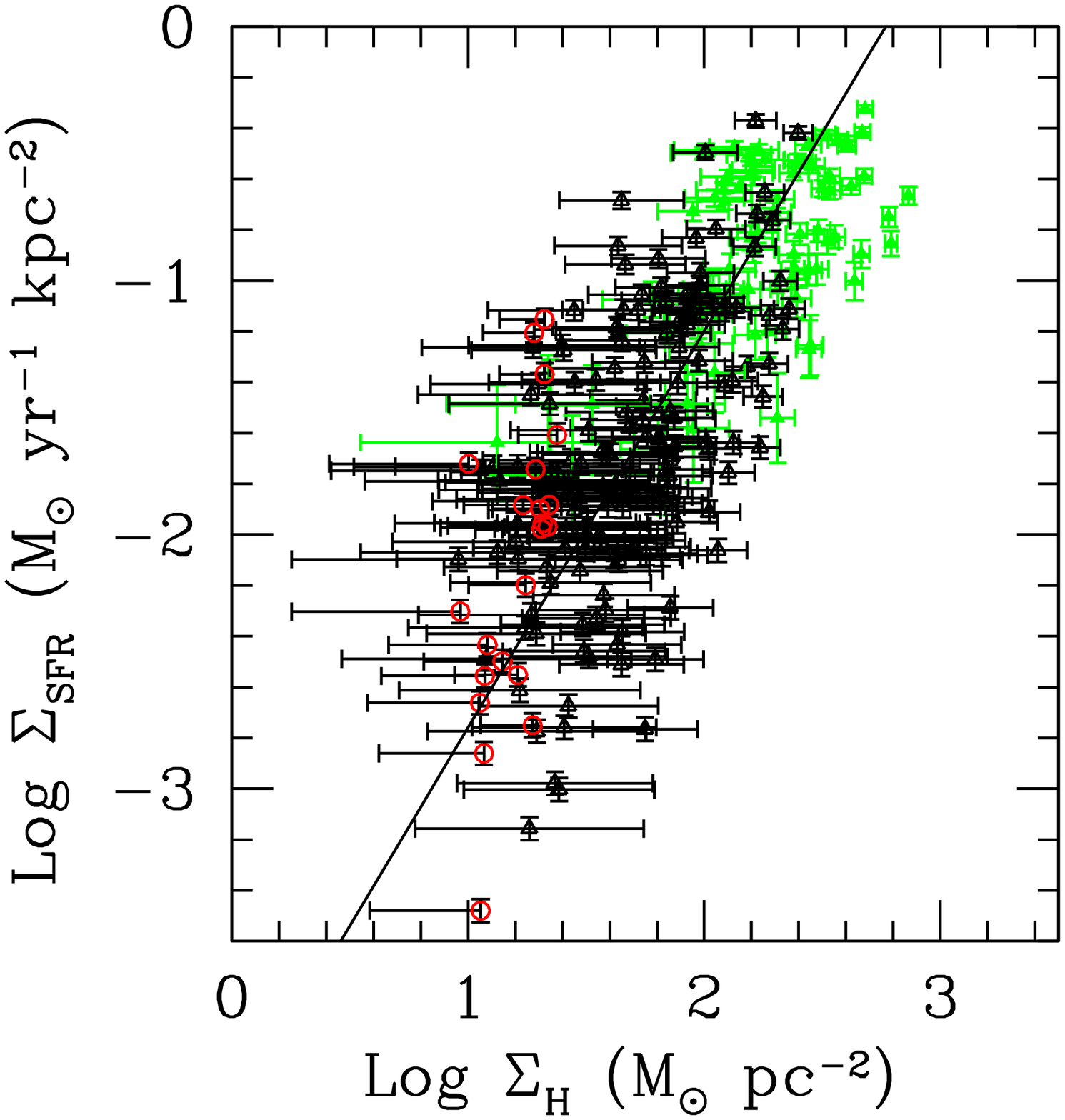}{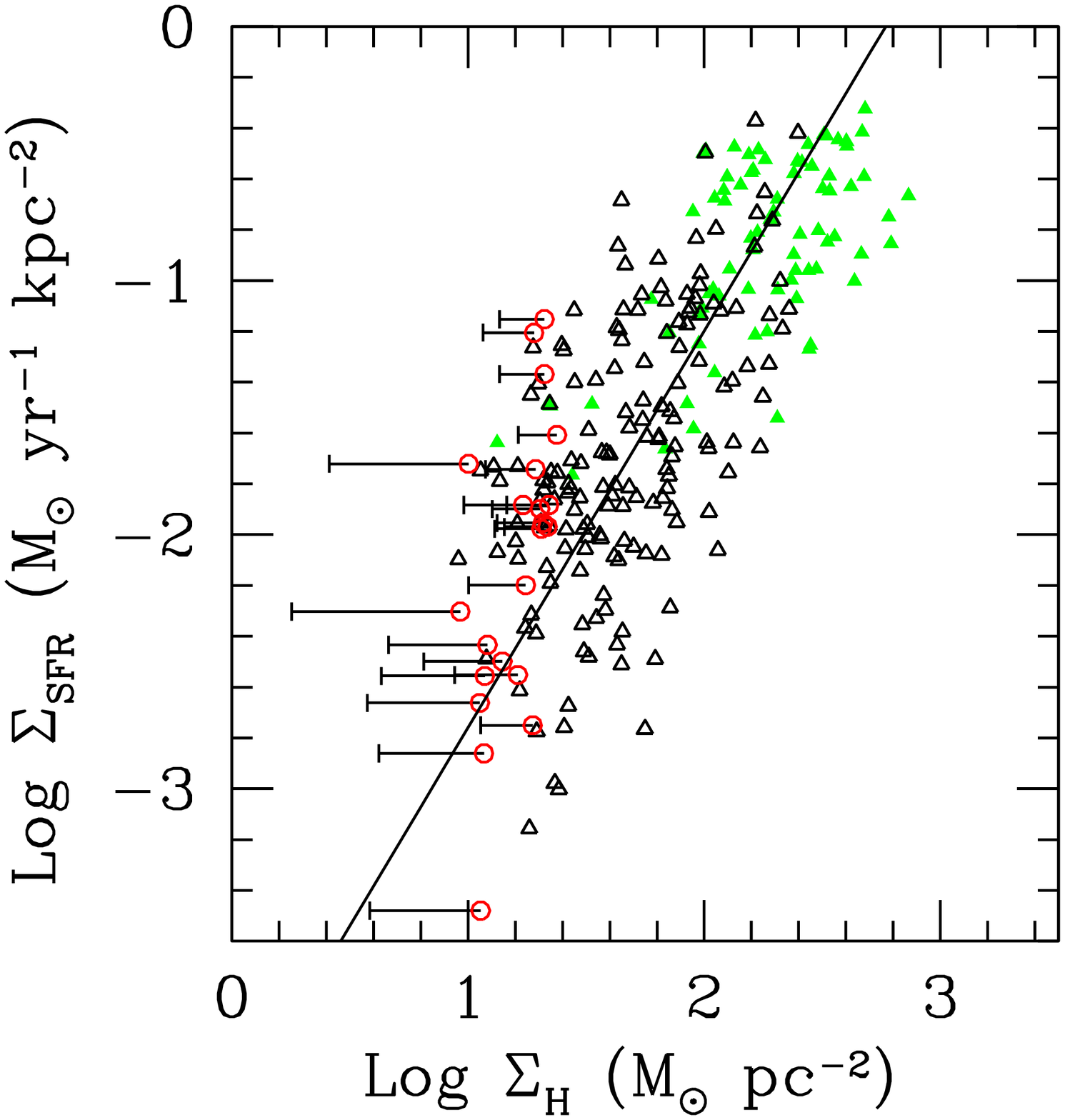}
\caption{Relationship between SFR surface
density and total (atomic plus molecular) hydrogen surface density for 257
HII regions and infrared sources measured in the central 350\arcsec\
region of M51 (the area covered in CO by BIMA SONG).  Solid green triangles
denote SFRs derived from extinction-corrected Pa$\alpha$ fluxes, all
in the central 144\arcsec, while open black triangles denote SFRs determined
from combined 24~\micron\ and H$\alpha$ fluxes, using the method
described in \S4.  Open (red) circles denote regions with only 3-$\sigma$ 
upper limits in CO (see text).} The fluxes were measured with aperture 
diameters
of 13\arcsec\ (520 pc).  The line shows a best fitting power law with
slope $N$ = 1.56.  Points in the two panels are identical 
except for the inclusion of error bars, for the sake of clarity.
\end{figure}

\clearpage

\begin{figure}
\epsscale{0.9}
\plotone{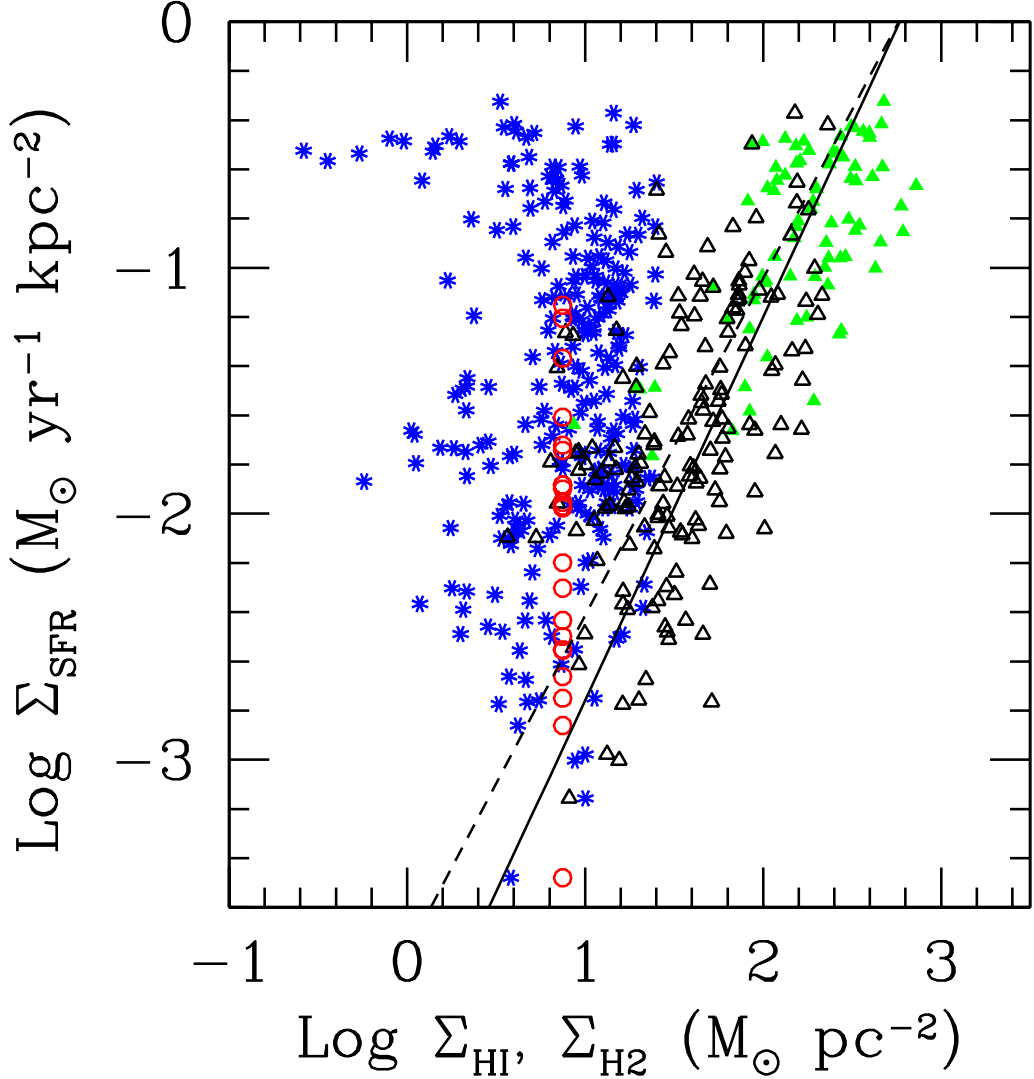}
\caption{Relation between local SFR density and molecular 
and atomic hydrogen surface densities separately.  The solid 
green and open black triangles denote H$_2$ surface densities (see
Fig.\ 4), with open red
circles indicating CO upper limits (same symbol notation
as for Figure 4).  Blue asterisks show the corresponding
relation between SFR surface densities and HI surface densities.  The
dashed line shows the best bivariate least squares fit
to the molecular densities alone.  The fit to total gas density
(see Figure 4) is shown for reference as the solid line.} 
\end{figure}

\clearpage

\begin{figure}
\epsscale{0.9}
\plotone{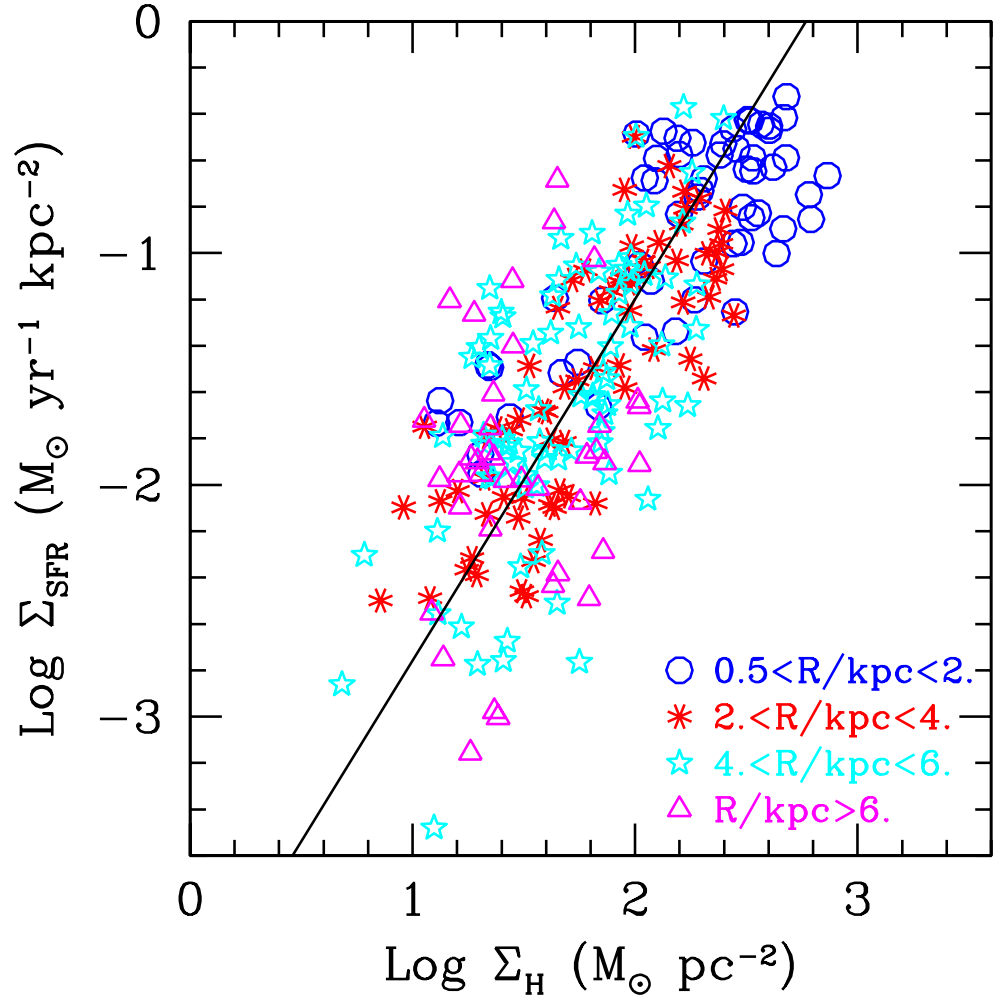} 
\caption{Same data as shown in Figure 4, but with points coded by 
galactocentric radius.}
\end{figure}

\clearpage

\begin{figure}
\epsscale{0.9}
\plotone{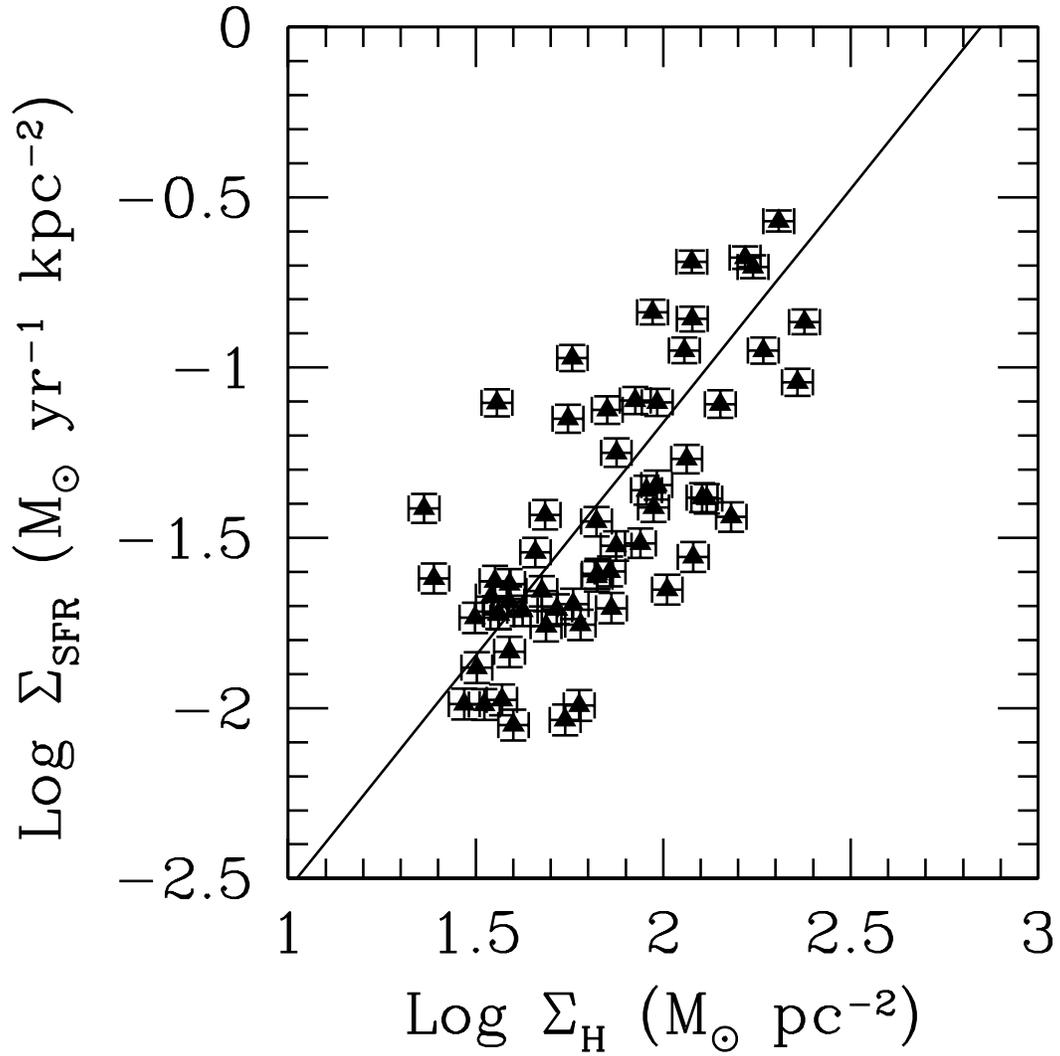}
\caption{Relation between SFR and total gas surface densities using CO 
and HI data from Lord \& Young (1990), with aperture diameters of 45\arcsec\ 
(1850 pc).  The solid line shows a best bivariate fit with slope 
$N$ = 1.37.}  
\end{figure}

\clearpage

\begin{figure}
\epsscale{0.9}
\plotone{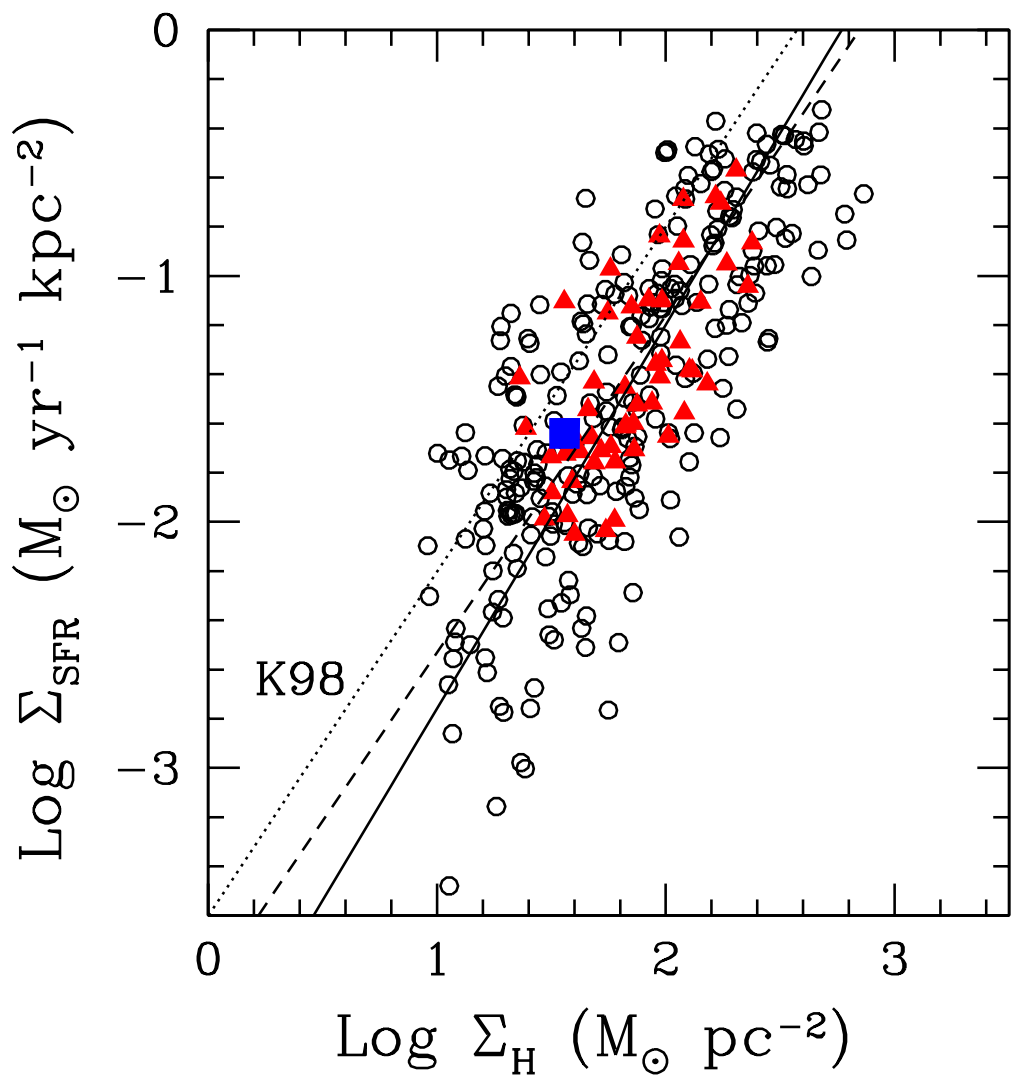}
\caption{Comparison of the Schmidt law measured for M51 using 520 pc 
(13\arcsec) apertures (open black circles) and 1850 pc (45\arcsec) 
apertures (solid red triangles), 
with the best fitting power-law fits shown with solid and dashed lines,
respectively.  Shown for comparison by the upper dotted line is the 
disk-averaged Schmidt law for normal and starburst galaxies from 
Kennicutt (1998b).  The large blue square shows the disk-averaged 
SFR and gas density for M51 as measured in that paper.}
\end{figure}

\clearpage

\begin{figure}
\epsscale{0.9}
\plotone{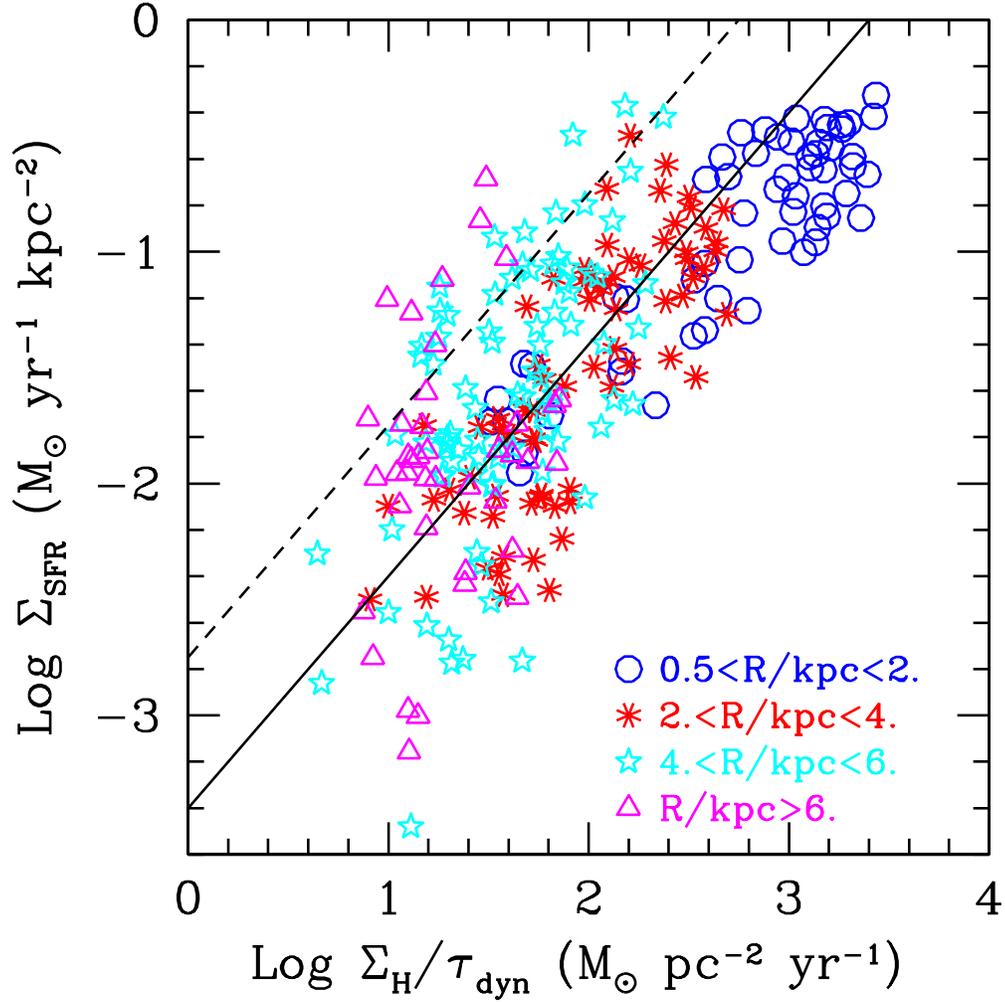}
\caption{SFR surface densities plotted as a function of the ratios of 
gas surface density
to orbit time (equation [2]).  The 520 pc aperture data are shown, and are 
color coded by galactocentric radius in kpc, 
as in Figure 6.  The solid line is the best fit relation with slope constrained
to unity.  Note the deviation from linear slope at constant radius, 
though the large-scale distribution of points traces a roughly linear
relation.  The dashed line shows the comparable fit to the disk-averaged SFRs
in Kennicutt (1998b).}
\end{figure}

\clearpage

\begin{figure}
\epsscale{0.9}
\plotone{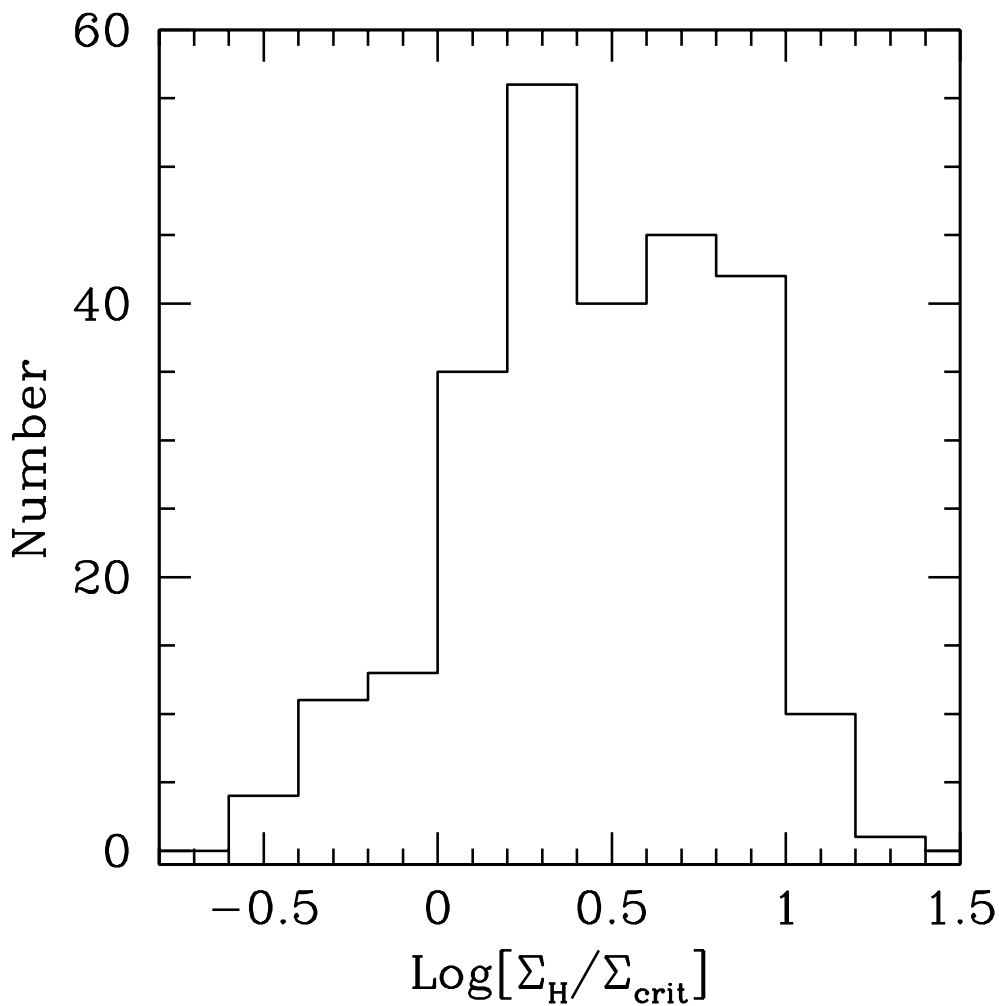}
\caption{Distribution of the total gas surface densities of the 257
regions in M51 studied here (520 pc apertures), normalized in each
case to the local critical density for gravitational stability,
as defined in eq. (9).  Expressed in terms of the Toomre stability
parameter $Q$ the ratio plotted corresponds roughly to $1/Q$.  
The densities include molecular and atomic
hydrogen, multiplied by a factor of 1.38 to account for helium and
metals.  Note the truncation near a value of SFR densities for 
$Q < 1$.}
\end{figure}

\end{document}